\documentclass[journal,onecolumn,11pt, letter]{IEEEtran}
\usepackage[doublespacing]{setspace}
\usepackage{fullpage}
\usepackage{commands11}
\usepackage{cite}
\usepackage{color}
\usepackage{graphicx}

\usepackage{amsfonts}
\usepackage{amssymb}
\usepackage{mathrsfs}

\usepackage{array}
\usepackage{paralist}
\usepackage{verbatim}
\usepackage{enumerate}
\usepackage{stfloats}

\newtheorem{theorem}{Theorem}
\newtheorem{lemma}{Lemma}
\newtheorem{corollary}{Corollary}
\newtheorem{definition}{Definition}
\newtheorem{proposition}{Proposition}

\begin{document}

\title{Capacity of Gaussian Channels with Duty Cycle and Power
  Constraints
  % \begin{comment}
  %   \thanks{This work was initiated during the authors' visit to the
  %     Institute of Network Coding at the Chinese University of Hong
  %     Kong.  The work of D.~Guo was partially supported by a grant
  %     from the University Grants Committee of the Hong Kong Special
  %     Administrative Region, China (Project No.~AoE/E-02/08). This
  %     work was also partially supported by the NSF under Grant
  %     No.~0644344 and by DARPA under Grant No.~W911NF-07-1-0028.}
  % \end{comment}
}

\author{Lei~Zhang$^\dag$, Hui~Li$^{\dag*}$, and~Dongning~Guo$^\dag$
  \\\vspace{1ex} \small \dag { Department of Electrical Engineering
    and Computer Science,
    Northwestern University, \\
    Evanston, IL 60208, USA}\\\vspace{1ex} * {Department of Electronic
    Engineering and Information Science,
    University of Science and Technology of China,\\
    Hefei, Anhui 230027, China} \thanks{ This work has been presented in part at the 2011 and 2012 IEEE International
    Symposium on Information Theory.}
% \IEEEmembership{Member,~IEEE,}
}

\IEEEcompsoctitleabstractindextext{
  \begin{abstract}
    In many wireless communication systems, radios are subject to a {\em
      duty cycle} constraint, that is, a radio only actively transmits
    signals over a fraction of the time.  For example, it is desirable
    to have a small duty cycle in some low power systems; a
    half-duplex radio cannot keep transmitting if it wishes to receive
    useful signals; and a cognitive radio needs to listen and detect
    primary users frequently.
    This work studies the capacity of scalar discrete-time Gaussian
    channels subject to duty cycle constraint as well as average
    transmit power constraint.  An idealized duty cycle constraint is
    first studied, which can be regarded as a requirement on the
    minimum fraction of nontransmissions or zero symbols in each
    codeword.  A unique {\em discrete} input distribution is shown to
    achieve the channel capacity.
    In many situations, numerically optimized on-off signaling can
    achieve much higher rate than Gaussian signaling over a
    deterministic transmission schedule. This is in part because the
    positions of nontransmissions in a codeword can convey
    information.
    Furthermore, a more realistic duty cycle constraint is studied,
    where the extra cost of transitions between transmissions and
    nontransmissions due to pulse shaping is accounted for. The
    capacity-achieving input is no longer independent over time and is
    hard to compute. A lower bound of the achievable rate as a
    function of the input distribution is shown to be maximized by a
    first-order Markov input process, the distribution of which is
    also {\em discrete} and can be computed efficiently.  The results
    in this paper suggest that, under various duty cycle constraints,
    departing from the usual paradigm of intermittent packet
    transmissions may yield substantial gain.

  \end{abstract}

  \clearpage
  \begin{IEEEkeywords}
    Duty cycle constraint, capacity-achieving input, mutual
    information, entropy rate, Markov process, hidden Markov
    process (HMP), Monte Carlo method.
  \end{IEEEkeywords}
}
\maketitle

\IEEEdisplaynotcompsoctitleabstractindextext \IEEEpeerreviewmaketitle

\section{Introduction}

In many wireless communication systems, a radio is designed to
transmit actively only for a fraction of the time, which is known as
its {\em duty cycle}.  For example, the ultra-wideband
system %proposed
in~\cite{JulMaj11ITA} %employs impulse radio to
transmits short bursts of signals to trade bandwidth for power
savings.  The physical half-duplex constraint also requires a radio to
stop transmission over a frequency band from time to time if it wishes
to receive useful signals over the same band.  Thus wireless relays
are subject to duty cycle constraint, so do cognitive radios which
have to listen to the channel frequently to avoid causing interference
to primary users. The {\em de facto} standard solution under duty
cycle constraint is to transmit packets intermittently.

This work studies the fundamental question of what is the optimal
signaling for a Gaussian channel with duty cycle constraint as well as
average transmission power constraint.  An important observation is
that the signaling in nontransmission periods can be regarded as
transmission of a special {\em zero} signal.  We first make a
simplistic and idealized assumption that the analog waveform
corresponding to each transmitted symbol is exactly of the span of one
symbol interval.  We restrict our attention to discrete-time scalar
additive white Gaussian noise (AWGN) channels for simplicity, where
the duty cycle constraint is equivalent to a requirement on the
minimum fraction of zero symbols in each transmitted codeword, which
is called the {\em idealized duty cycle constraint}. We then consider the
case where a practical pulse shaping filter is used, e.g., for
band-limited transmissions. As such, during a transition between a
zero symbol and a nonzero symbol, the pulse waveform of the nonzero
symbol leaks into the interval of the zero symbol. A realistic duty
cycle constraint must include the extra cost incurred upon transitions
between zero and nonzero symbols.  The mathematical model of the preceding input-constrained channels is described in Section~\ref{sec:Model}.

% In order to alleviate such impact in practice, designs for pulse
% shaping filters need to be taken into consideration. In this work,
% however, we restrict our focus on the discrete-time model.

Determining the capacity of a channel subject to various input
constraints is a classical problem. It is well-known that Gaussian
signaling achieves the capacity of a Gaussian channel with average
input power constraint only. In addition, Zamir\cite{Zamir2004IT}
shows that the mutual information rate achievable using a white Gaussian input never
incurs a loss of more than half a bit per sample with respect to the
power constrained capacity.
Furthermore, Smith~\cite{Smith71} investigated the
capacity of a scalar AWGN channel under both peak power constraint and
average power constraint. The input distribution that achieves the
capacity is shown to be discrete with a finite number of probability
mass points. The discreteness of capacity-achieving distributions for
various channels, including quadrature Gaussian channels, and
Rayleigh-fading channels is also established in~\cite{Shamai90,
  ShaBar95IT, AboTro01IT, KatSha04IT, GurPoo05WC,
  HuaMey05IT}. Chan~\cite{ChaHra05IT} studied the capacity-achieving
input distribution for conditional Gaussian channels which form a
general channel model for many practical communication
systems.
%\marginpar{\scriptsize  \parbox[t]{0.6in}{to be deleted upon acceptance}}
Until now, the impact of duty cycle constraint on capacity-achieving signaling is underexplored in the literature.

The main results of this paper are summarized in
Section~\ref{sec:main-results}. In the case of the idealized duty
cycle constraint, because all costs associated with the constraints
can be decomposed into per-letter costs, the optimal input
distribution is independent and identically distributed (i.i.d.). We
use a similar approach as in~\cite{Smith71} and~\cite{ChaHra05IT} to
show that the capacity-achieving input distribution for an AWGN
channel with duty cycle constraint and average power constraints is
discrete. Unlike in ~\cite{Smith71} and~\cite{ChaHra05IT}, the optimal
distribution has an infinite number of probability mass points,
whereas only a finite number of the points are found in every bounded
interval. This allows efficient numerical optimization of the input
distribution.

The case of realistic duty cycle constraint is more
challenging. Because the constraint concerns symbol transitions, the
capacity-achieving input distribution is no longer independent over
time, and becomes hard to compute. We develop a good lower bound of the
input-output mutual information as a function of the input
distribution. It is proved that, under the realistic duty cycle
constraint, a first-order Markov process maximizes the lower bound,
the distribution of which is also discrete and can be computed
efficiently. The main theorems for the cases of idealized and
realistic duty cycle constraints are proved in Section
\ref{sec:MainRes} and \ref{sec:lower-bound-capacity}, respectively.

We devote Section \ref{sec:numer-meth-results} to the numerical
methods and results. In order to compute the achievable rate when the
input is a Markov Chain, a Monte Carlo method is introduced in Section
\ref{sec:numer-comp-entr} to numerically compute the differential
entropy rate of hidden Markov processes.  Numerical results in
Section~\ref{sec:NumRes} demonstrate that in the case of idealize duty
cycle constraint using a numerically optimized discrete signaling
achieves higher rates
% with finite probability mass points compared to using
than using Gaussian signaling over a deterministic transmission
schedule.  For example, if the radio is allowed to transmit no
more than half the time, i.e., the duty cycle is no greater than 50\%,
a near-optimal discrete input achieves 50\% higher rate at 10~dB
signal-to-noise ratio (SNR). In the case of realistic duty cycle
constraint, numerical results also show that the rate achieved by the
Markov process is substantially higher than that achieved by any
i.i.d. input. This suggests that, compared to intermittently
transmitting packets using Gaussian or Gaussian-like signaling, it is
more efficient to disperse nontransmission symbols within each packet
to form codewords, which results in a form of {\em on-off} signaling.

One of the reasons for the superiority of on-off signaling is that the
positions of nontransmission symbols can be used to convey
information, the impact of which is particularly significant in case
of low SNR or low duty cycle.  This has been observed in the past.
For example, as shown in~\cite{LutHau08ISIT} (see also~\cite{Kramer07,
  LutKra10ISIT}), time sharing or time-division duplex (TDD) can fall
considerably short of the theoretical limits in a relay network: The
capacity of a cascade of two noiseless binary bit pipes through a
half-duplex relay is 1.14 bits per channel use, which far exceeds the
0.5 bit achieved by TDD and even the 1 bit upper bound on the rate of
binary signaling.

Besides that duty cycle constraint is frequently seen in practice,
another motivation of this study is a recent
work~\cite{GuoZha10Allerton}, in which on-off signaling is proposed
for a clean-slate design of wireless ad hoc networks formed by
half-duplex radios.  Using this signaling scheme, which is called
rapid on-off-division duplex (RODD), a node listens to the channel and
receives useful signals during its own off symbols within each frame.
Each node can transmit and receive messages at the same time over one
frame interval, thereby achieving (virtual) full-duplex
communication. Understanding the impact of duty cycle constraint is
crucial to characterizing the fundamental limits of such wireless
networks.

\section{System Model} \label{sec:Model}

Consider digital communication systems where coded data are mapped to
waveforms for transmission.  Usually there is a collection of pulse
waveforms, where each pulse represents a symbol (or letter) from a
discrete alphabet.  We view nontransmission over a symbol interval as
transmitting the all zero waveform.  In other words, a symbol interval
of nontransmission is simply regarded as transmitting a special symbol ``0,'' which carries no energy.

As far as the capacity-achieving input is concerned it suffices to
consider the baseband discrete-time model for the AWGN channel. The
received signal over a block of $n$ symbols can be described by
\begin{align} \label{eq:Y=X+N} Y_i=X_i+N_i
\end{align}
where $i=1,\dots,n$, $X_i$ denotes the transmitted symbol at time $i$
and $N_1,\dots,N_n$ are independent standard Gaussian random
variables. For simplicity, we assume no inter-symbol interference is
at receiver.
Each symbol modulates a continuous-time pulse waveform for
transmission. If the width of all pulses were exactly of one symbol
interval, which is denoted by $T$, the duty cycle is equal to the
fraction of nonzero symbols in a codeword. In practice, however, the
pulse is usually wider than $T$, so that the support of the
transmitted waveform is greater than the sum of the intervals
corresponding to nonzero symbols due to leakage into intervals of
adjacent zero symbols. To be specific, suppose the width of a pulse is
$(1+2c)T$, then each transition between zero and nonzero symbols
incurs an additional cost of up to $c\,T$ in terms of actual
transmission time.

Let $1-q$ denote the maximum duty cycle allowed. In this paper, we
require every codeword $(x_1, x_2, \cdots, x_n)$ to satisfy
\begin{align}
  \label{eq:DC_cons}
  \frac{1}{n}\sum_{i=1}^n
  1_{\{x_i\ne0\}}+\frac{1}{n}2c\left(\sum_{i=1}^{n-1}1_{\{x_i=0, x_{i+1}\ne0\}}+1_{\{x_n=0,x_1\ne0\}}\right)\le 1-q
\end{align}
where $1_{\{\cdot\}}$ is the indicator function, and the transition
cost is twice that of zero-to-nonzero transitions, because the number
of nonzero-to-zero transitions and the number of zero-to-nonzero
transitions is equal under the cyclic transition cost
configuration. From now on, we refer to (\ref{eq:DC_cons}) as {\em
  duty cycle constraint $(q,c)$}. Note that the idealized duty cycle
constraint is the special case $(q,0)$.  If $c\in[0,\frac{1}{2}]$,
then the left hand side of (\ref{eq:DC_cons}) is equal to the actual
duty cycle. If $c>\frac{1}{2}$, the left hand side of
(\ref{eq:DC_cons}) is an overestimate of the duty cycle. Nonetheless,
we use constraint (\ref{eq:DC_cons}) for its simplicity.
In addition, we consider the usual average input power constraint,
\begin{align}
\frac{1}{n}\sum_{i=1}^n x_i^2\le \gamma.\label{eq:31}
\end{align}

In many wireless systems, the transmitter's activity is constrained in
the frequency domain as well as in the time domain.  In principle, the
results in this paper also apply to the more general model where the duty cycle
constraint is on the time-frequency plane.
% It may be harder to implement sharp band-pass filters to remove
% self-interference.

\section{Main Results}\label{sec:main-results}

\subsection{The Case of Idealized Duty Cycle Constraint}

Let $\mu$ denote the distribution of the channel input $X$.   The set of
distributions with duty cycle constraint $(q,0)$ and power constraint
$\gamma$ is denoted by
\begin{align}
  \label{eq:4}
  \Lambda(\gamma,q)&=\big\{\mu:\mu(\{0\}) \geq q, \,
  \expsub{\mu}{X^2}\leq \gamma\}.
\end{align}
It should be understood that $\mu$ is a probability measure defined on
the Borel algebra on the real number set, denoted by
$\mathcal{B}(\reals)$.

\begin{theorem} \label{thm:Main}
  The capacity of the additive white Gaussian noise
  channel~\eqref{eq:Y=X+N} with its idealized duty cycle no greater
  than $1-q$ and the average power no greater than $\gamma$ is
  \begin{align}
    \label{eq:Capacity1}
    C(\gamma,q) = \max_{\mu\in\Lambda(\gamma,q)}  I(\mu)\,.
  \end{align}
  In particular, the following properties hold:
  \begin{compactenum}
  \item[a)] the maximum
    of~\eqref{eq:Capacity1} % the mutual information
    is achieved by a unique (capacity-achieving) distribution $\mu_0
    \in \Lambda(\gamma,q)$;
  \item[b)] $\mu_0$ is symmetric about 0 and its second moment is
    exactly equal to $\gamma$; and
  \item[c)] $\mu_0$ is discrete with an infinite number of probability
    mass points, whereas the number of probability mass points in any
    bounded interval is finite.
  \end{compactenum}
\end{theorem}

The proof of Theorem \ref{thm:Main} is relegated to Section
\ref{sec:MainRes}. Property (b) suggests that the capacity-achieving
input always exhausts the power budget. Property (c) indicates that the
capacity-achieving input can be well approximated by some discrete
inputs with finite alphabet, which can be computed using numerical
methods. The achievable rate of numerically optimized input
distribution is studied in Section \ref{sec:numer-meth-results}.

\subsection{The Case of Realistic Duty Cycle Constraint}
In this paper, let $X_k^n$ denote the subsequence $(X_k,X_{k+1},
\cdots, X_n)$, where $X_k^\infty=(X_k,X_{k+1},\cdots)$. We also use
shorthand $X^n=X_1^n$.
Let $\mu$ denote the probability distribution of the process
$X_1,X_2,\cdots$. We use $\mu_{X_i}$ to denote the marginal
distribution of $X_i$, and $\mu_{X_{i},X_j}$ to denote the joint
probability distribution of $(X_i,X_j)$.
Denote the set of $n$-dimension distribution which satisfy duty cycle
constraint $(q,c)$ and power constraint $\gamma$ by
\begin{align}
  \Lambda^n(\gamma,q,c)= \Bigg\{ \mu:\;
  & \frac1n \sum^n_{i=1} \left[ \mu_{X_i}(\{0\})-2\,c\,\mu_{X_i,X_{i\text{
          mod }n+1}}(\{0\}\times(\reals\backslash\{0\}))
  \right] \ge q, \nonumber\\
  & \expsub{\mu}{\frac1n\sum^n_{i=1} X_i^2}\le \gamma \Bigg\}\label{eq:28}
\end{align}
where
\begin{align}
  \mu_{X_i,X_j}(\{0\}\times(\reals\backslash\{0\})) = P(X_i=0,X_j\ne0)
\end{align}
denotes the probability of a zero-to-nonzero transition and
\begin{align}
  i \text{ mod } n =
  \begin{cases}
    i, \quad &\text{if } 1\le i<n,\\
    0,\quad &\text{if } i=n.
  \end{cases}
\end{align}
For convenience in a subsequent proof, the duty cycle in~\eqref{eq:28}
is defined in a cyclic manner using the modular operation, where a
transition between $X_n$ and $X_1$ is also counted.  This of course
has vanishing impact as $n\to\infty$ and thus no impact on the capacity.

% \begin{align}
%   \Lambda^n(P,q,c)= \big\{ \mu:\;
%   &\mu_{X_i}(\{0\})-2c\mu_{X_i,X_{i+1}}(\{x_i=0,x_{i+1}\ne0\})\ge q,
%   \;i=1,\cdots,n-1; \nonumber\\
%   &\mu_{X_n}(\{0\})-2c\mu_{X_n,X_{1}}(\{x_n=0,x_{1}\ne0\})\ge q;
%   \nonumber \\
%   & \expsub{\mu}{X_i^2}\le P,\;i=1,\cdots,n\big\}.\label{eq:28}
% \end{align}

% Let $\Theta(P,q,c)$ denote the the subset of $\Lambda(P,q,c)$ with the
% average power equal to $P$, where
% \begin{equation}
%   \Theta(P,q,c)=\Lambda(P,q,c)\cap
%   \{\mu:\mathsf{E}_\mu\,\{X^2\}=P\}.\label{eq:19n}
% \end{equation}
The capacity of the AWGN channel (\ref{eq:Y=X+N}) with
duty cycle constraint $(q,c)$  and %average
power constraint $\gamma$ is
\begin{align}
  \label{eq:11}
  \begin{split}
    C(\gamma,q,c)&= \lim_{n\to\infty} \frac{1}{n} \; \max_{P_{X^n} \in
      \Lambda^n(\gamma,q,c)} I(X^n;Y^n).
  \end{split}
\end{align}

The capacity is in fact achieved by a stationary input process.  This is
justified in Section~\ref{sec:cons-about-stat} by showing that any
nonstationary input process has a stationary counterpart with equal or
greater input-output mutual information per symbol.
Let us denote the set of stationary distributions which satisfy duty
cycle constraint $(q,c)$ and power constraint $\gamma$ by
\begin{align}
  \begin{split}
    \Lambda(\gamma,q,c)= \big\{ \mu:\;& \mbox{$\mu$ is stationary, }\;
    \expsub{\mu}{X_1^2}\le \gamma, \\
    &\mu_{X_1}(\{0\})-2\,c\,\mu_{X_1,X_2}(\{0\}\times(\reals\backslash\{0\}))\ge
    q\big\}. \label{eq:9n}
  \end{split}
\end{align}

\begin{theorem}\label{thm:lowerbound}
  For any $\mu \in \Lambda(\gamma,q,c)$, let
  \begin{align}
    \label{eq:21n}
    L(\mu)=I(X;Y)-I(X_1;X_2^\infty)
  \end{align}
  where $I(X;Y)$ is the mutual information of the additive white
  Gaussian noise channel between the input symbol $X$, which follows
  distribution $\mu_{X_1}$, and the corresponding output $Y$. The
  following properties hold:
  \begin{compactenum}
  \item[a)] $L(\mu)$ is a lower bound of the channel capacity;
    % \begin{align}
    %   \label{eq:22n}
    %   I(\mu) \ge I(X;Y)-I(X_1;X_2^\infty).
    % \end{align}
    \label{enm:lb-lowerbound}
  \item[b)] The maximum of $L(\cdot)$ is achieved by a discrete first-order Markov process, denoted by $\mu^\ast$; \label{enm:lb-markov}
  \item[c)] $\mu^\ast$ satisfies the following property: Define $B_i=1_{\{X_i\ne0\}},i=1,2,\dots$. Then for every $i$, conditioned on $B_i$ and $B_{i+1}$, the variables $X_i$ and $X_{i+1}$ are independent, and
    \begin{align}
      \label{eq:1}
      L(\mu^*)=I(X;Y)-I(B_1;B_2).
    \end{align}\label{enm:lb-Imu}
  \end{compactenum}
\end{theorem}

The proof of Theorem \ref{thm:lowerbound} is relegated to Section
\ref{sec:lower-bound-capacity}. Evidently, increasing the input power
by scaling the input linearly not only maintains its duty cycle, but
also increases the mutual information. Therefore, the optimal input
distribution must exhaust the power budget $\gamma$.

\section{Proof of Theorem \ref{thm:Main} (the Case of Idealized Duty
  Cycle Constraint)} \label{sec:MainRes}

\begin{comment}
  Consider the additive white Gaussian noise channel described by the
  following relationship between its input $X$ and output $Y$:
  \begin{align} \label{eq:Y=X+N} Y= X+ N
  \end{align}
  where $N$ is standard Gaussian with probability density function
  (pdf)
  \begin{align} \label{eq:3} \phi(t) = \frac1{\sqrt{2\pi}} e^{-t^2/2}
    \, .
  \end{align}
\end{comment}

This section is devoted to a proof of Theorem~\ref{thm:Main} for the case of the idealized duty cycle constraint $(q,0)$. The conditional probability density function (pdf) of the output given the input of the AWGN channel~\eqref{eq:Y=X+N} is
\begin{align}
  \label{eq:2}
  p_{Y|X}(y|x) = \phi(y-x)
\end{align}
where
\begin{align} \label{eq:3}
\phi(t) = \frac1{\sqrt{2\pi}} e^{-\frac{t^2}{2}}
\end{align}
is the standard Gaussian pdf.

With the idealized constraint, the capacity of the AWGN
channel is achieved by an i.i.d. process and the duty cycle constraint
reduces to a per symbol cost constraint.  For given input distribution $\mu$, the pdf of the output exists and is expressed as
\begin{align} \label{eq:pY} p_Y(y;\mu) = \int p_{Y|X}(y|x) \,\mu(\diff
  x) = \expsub{\mu}{ \phi(y-X) } \,.
\end{align}
Denote the relative entropy
$D\left(p_{Y|X}(\cdot|x)\|p_Y(\cdot;\mu)\right)$ by $d(x;\mu)$, which
is expressed as
\begin{align} \label{eq:dxmu} d(x;\mu)=\int_{-\infty}^{\infty}
  p_{Y|X}(y|x)\log\frac{p_{Y|X}(y|x)}{p_Y(y;\mu)} \diff y \ .
\end{align}
The mutual information $I(\mu)=I(X;Y)$ is then
\begin{align} \label{eq:9} I(\mu)=\int d(x;\mu) \,\mu(\diff x)
  =\expsub{\mu}{d(X;\mu)}.
\end{align}

The capacity of the AWGN channel under per-letter duty cycle constraint and power constraint is evidently given by the supremum of the mutual information $I(\mu)$ where $\mu\in\Lambda(\gamma,q)$. The achievability and converse of this result can be established using standard techniques in information theory.

The proof of property~(a) is presented in Section~\ref{sec:propa}. Now suppose $\mu_0$ is the unique capacity-achieving distribution, property~(b) is established as follows.  Since the mirror reflection of $\mu_0$ about 0 is evidently also a maximizer of~\eqref{eq:Capacity1}, the uniqueness requires that $\mu_0$ be symmetric.  Note that linear scaling of the input to increase its power maintains its duty cycle and cannot reduce the mutual information, as the receiver can add noise to maintain the same SNR. By the uniqueness of the maximizer $\mu_0$, the power constraint must be binding, i.e., the second moment of $\mu_0$ must be equal to $\gamma$. In order to prove property~(c), we first establish a sufficient and necessary condition for $\mu_0$ in Section~\ref{sec:sn} and then apply it to show the discreteness of $\mu_0$ in Section~\ref{sec:propc}.

\subsection{Existence and Uniqueness of $\mu_0$}
\label{sec:propa}

Let $\mathcal{P}$ denote the collection of all Borel probability measures defined on $(\reals,\mathcal{B}(\reals))$, which is a topological space with the topology of weak convergence~\cite{Stroock93}. We first establish the following lemma.
\begin{lemma}\label{lm:compact}
    $\Lambda(\gamma,q)$ is compact in the topological space $\mathcal{P}$.
\end{lemma}
\begin{IEEEproof}
According to~\cite{Stroock93}, the topology of weak convergence on $\mathcal{P}$ is metrizable.
%whose metric is called the L\'{e}vy-Prohorov metric.
Therefore, by Prokhorov's theorem~\cite{Pro56TPA}, in order to prove that $\Lambda(\gamma,q)$ is compact in $\mathcal{P}$, it suffices to show that it is both tight and closed.

For any $\epsilon>0$, there exits an $a_{\epsilon}>0$, such that for all $\mu \in \Lambda_\gamma$,
\begin{align}
\mu(|X|> a_{\epsilon})\leq \frac{\expsub{\mu}{X^2}}{a_{\epsilon}^2}\leq \frac{\gamma}{a_{\epsilon}^2}<\epsilon
\end{align}
by Chebyshev's inequality. Choose $K_{\epsilon}=[-a_{\epsilon},a_{\epsilon}]$, then $K_{\epsilon}$ is compact in $\reals$ and $\mu(K_{\epsilon})\geq 1-\epsilon$ for all $\mu \in \Lambda(\gamma,q)$, thus $\Lambda(\gamma,q)$ is tight.

Let $B_m=\left[-\frac{1}{m},\frac{1}{m}\right]$ for $m=1,2,\dots$. Let $\{\mu_n\}_{n=1}^{\infty}$ be a convergent sequence in $\Lambda(\gamma,q)$ with limit $\mu_0$. Since $\mu_n(B_m)\geq q$ for every $m,n$, we have~\cite[Section~$3.1$]{Stroock93}
\begin{align}
q\leq\limsup_{n\rightarrow\infty}\mu_n(B_m)\leq \mu_0(B_m),
\end{align}
and hence
\begin{align}
\mu_0(\{0\})=\mu_0\left(\bigcap_{m=1}^{\infty}B_m\right)=\lim_{m\rightarrow\infty}\mu_0(B_m)\geq q.
\end{align}
Moreover, let $f(x)=x^2$ which is continuous and bounded below. By weak convergence~\cite[Section~$3.1$]{Stroock93}, we have
\begin{align}
\expsub{\mu_0}{X^2}=\int f\diff\mu_0\leq \liminf_{n\rightarrow\infty} \int f\diff\mu_n \leq \gamma.
\end{align}
Therefore, $\mu_0 \in \Lambda(\gamma,q)$, i.e., $\Lambda(\gamma,q)$ is closed, and the compactness of $\Lambda(\gamma,q)$ then follows.
\end{IEEEproof}

Since the mutual information $I(\mu)$ is continuous on $\mathcal{P}$~\cite[Theorem~$9$]{WuVer10ISIT}, it must achieve its maximum on the compact set $\Lambda(\gamma,q)$. Hence the capacity-achieving distribution $\mu_0$ exists.

According to~\cite[Corollary~2]{WuVer10ISIT}, the mutual information
$I(\mu)$ is strictly concave.  It is easy to see that $\Lambda(\gamma,q)$ is
convex.  Hence the capacity-achieving distribution $\mu_0$ must be unique.

\subsection{Sufficient and Necessary Conditions}
\label{sec:sn}

We denote the finite-power set as
\begin{align}
\Lambda(q)= \cup_{0\le \gamma<\infty} \Lambda(\gamma,q).
\end{align}
Let $\phi(\cdot)$ defined in~\eqref{eq:3} be extended to the complex plane. The relative entropy $d(x;\mu)$ defined in~\eqref{eq:dxmu} can be extended to the complex plane $\complex$ and has the following property:
\begin{lemma} \label{lm:holo}
For any $\mu \in \Lambda(q)$ and $z \in \complex$,
\begin{align} \label{eq:hw} d(z;\mu)=\int_{-\infty}^{\infty}
  \phi(y-z) \log \frac{\phi(y-z)}{p_Y(y;\mu)} \diff y
\end{align}
is a holomorphic function of $z$ on $\complex$. Consequently, $d(x;\mu)$ is a continuous function of $x$ on $\reals$.
\end{lemma}
\begin{IEEEproof}
It can be shown that $\int_{-\infty}^{\infty} \phi(y-z) \log \phi(y-z) \diff y$ is a constant, thus a holomorphic function of $z$ on $\complex$.  Therefore, it remains to prove that
\begin{align} \label{eq:xi}
\xi(z)=\int_{-\infty}^{\infty} \phi(y-z) \log p_Y(y;\mu) \diff y
\end{align}
is a holomorphic function of $z$ on $\complex$.

First, by Jensen's inequality, we have
\begin{align}
p_Y(y;\mu) &= \expsub{\mu}{\frac{1}{\sqrt{2\pi}}e^{-\frac{(y-X)^2}{2}}} \\
&\geq \frac{1}{\sqrt{2\pi}}e^{-\frac{1}{2}\expsub{\mu}{(y-X)^2}} \\
&= e^{-\frac{1}{2}y^2-ay-b}
\end{align}
where $a=-\expsub{\mu}{X}$ and $b=\frac{1}{2}\left(\expsub{\mu}{X^2}+\log(2\pi)\right)$ are real numbers due to the fact that $\mu\in\Lambda(q)$. Thus, $p_Y(y;\mu) \in [e^{-\frac{1}{2}y^2-ay-b},1]$, i.e.,
\begin{align} \label{eq:logp_ub}
|\log P_Y(y;\mu)|\leq \frac{1}{2}y^2+ay+b.
\end{align}
As a result, we have
\begin{comment}
\begin{align}\label{eq:h_UB}
&\int_{-\infty}^\infty \left|\phi(y-z)\log p_Y(y;\mu)\right|\diff y \\
& \leq \int_{-\infty}^\infty \frac{1}{\sqrt{2\pi}}\left|e^{-\frac{(y-z)^2}{2}}\right| \left(\frac{1}{2}y^2+ay+b\right) \diff y \\
& = \frac{1}{\sqrt{2\pi}}e^{\frac{\text{Im}^2(z)}{2}} \int_{-\infty}^\infty e^{-\frac{(y-\text{Re}(z))^2}{2}}\left(\frac{1}{2}y^2+ay+b\right)\diff y < \infty
\end{align}
\end{comment}
\begin{align}\label{eq:h_UB}
\left|\phi(y-z)\log p_Y(y;\mu)\right|
& \leq \frac{1}{\sqrt{2\pi}}\left|e^{-\frac{(y-z)^2}{2}}\right| \left(\frac{1}{2}y^2+ay+b\right) \\
& = \frac{1}{\sqrt{2\pi}} e^{-\frac{(y-\text{Re}(z))^2-\text{Im}^2(z)}{2}}\left(\frac{1}{2}y^2+ay+b\right),
\end{align}
which is integrable. (Here $\text{Re}(z)$ and $\text{Im}(z)$ represent the real and imaginary parts of $z$, respectively.) It follows that $\xi(z)$ given by~\eqref{eq:xi} exists for any $\mu\in \Lambda(q)$ and $z\in\complex$.

Suppose $U$ is an open and bounded subset of $\complex$. There exists an $r>0$ such that $|\text{Re}(z)| \leq r$ and $|\text{Im}(z)| \leq r$ for all $z \in U$. It is easy to check that
\begin{align}
e^{-\frac{(y-\text{Re}(z))^2}{2}} &\leq e^{-\frac{y^2}{2}+|yr|} \\
&\leq e^{-\frac{y^2}{2}+yr}+e^{-\frac{y^2}{2}-yr} \\
&=e^{\frac{r^2}{2}}\left[e^{-\frac{1}{2}(y-r)^2}+e^{-\frac{1}{2}(y+r)^2}\right]. \label{eq:ub}
\end{align}
Combining~\eqref{eq:h_UB} and~\eqref{eq:ub} yields that
\begin{align}
\left|\phi(y-z)\log p_Y(y;\mu)\right|
&\leq \frac{e^{r^2}}{\sqrt{2\pi}}\left[e^{-\frac{1}{2}(y-r)^2}+e^{-\frac{1}{2}(y+r)^2}\right] \left(\frac{1}{2}y^2+ay+b\right),
\end{align}
which is integrable. Therefore, the integral $\int_{-\infty}^\infty \phi(y-z)\log p_Y(y;\mu) \diff y$ is uniformly convergent for all $z \in U$. Moreover, $\phi(y-z)\log p_Y(y;\mu)$ is a holomorphic function of $z$ on $U$ for each $y\in\reals$. According to the differentiation lemma~\cite{Lang99}, $\xi(z)$ is a holomorphic function of $z$ on $U$. It then follows that it is holomorphic on the whole complex plane $\complex$. Lemma~\ref{lm:holo} is thus established.
\end{IEEEproof}

Let $F(\mu)$ be a real-valued function defined on the convex set $\Lambda(q)$ and $\mu_0\in\Lambda(q)$. Define the weak derivative of $F(\mu)$ at $\mu_0$ as
\begin{align} \label{eq:Jwd}
F'_{\mu_0}(\mu) = \lim_{\theta\rightarrow 0^{+}}\frac{F\left((1-\theta)\mu_0+\theta\mu\right)-F(\mu_0)}{\theta}
\end{align}
whenever the limit exists.
%\marginpar{\scriptsize  \parbox[t]{0.6in}{to be deleted upon acceptance}}
The following result, which finds its parallel in~\cite{AboTro01IT,HuaMey05IT,ChaHra05IT} gives the weak derivative of the mutual information function $I(\mu)$.

\begin{lemma} \label{lm:wd}
Let $\mu_0, \mu\in\Lambda(q)$, the weak derivative of the mutual information function $I(\mu)$ at $\mu_0$ is
\begin{align}
I'_{\mu_0}(\mu)=\int d(x;\mu_0) \,\mu(\diff x)-I(\mu_0).
\end{align}
\end{lemma}
\begin{IEEEproof}
Define $\mu_\theta=(1-\theta)\mu_0+\theta\mu$ for all $\theta\in(0,1]$. It can be shown that
\begin{align}
\frac{1}{\theta}\left(I(\mu_\theta)-I(\mu_0)\right) =
&\,\frac{1}{\theta}\int \left(d(x;\mu_\theta)-d(x;\mu_0)\right)
\,\mu_\theta(\diff x) %\nonumber \\ &
+\frac{1}{\theta}\left(\int d(x;\mu_0) \,\mu_\theta(\diff x)-I(\mu_0)\right) \\
&= -\frac{1}{\theta}\int_{-\infty}^{\infty}
p_Y(y;\mu_\theta)\log\frac{p_Y(y;\mu_\theta)}{p_Y(y;\mu_0)}\diff y
%\nonumber \\&
+\int d(x;\mu_0) \,\mu(\diff x)-I(\mu_0).
\end{align}
Therefore, it suffices to show that
\begin{align} \label{eq:lim0}
\lim_{\theta\rightarrow0^{+}}\int_{-\infty}^{\infty} \frac{1}{\theta}p_Y(y;\mu_\theta)\log\frac{p_Y(y;\mu_\theta)}{p_Y(y;\mu_0)}\diff y=0.
\end{align}

In the remainder of this proof, we find a function independent of $\theta$ that dominates the integrand so that dominated convergence theorem can be used to establish~\eqref{eq:lim0} by exchanging the order of the limit and the integral therein.

\begin{lemma} \label{lm:ubd}
Let $\theta,a,b\in(0,1]$. Define
\begin{align}
f(\theta)=\frac{(1-\theta)a+\theta b}{\theta}\log\frac{(1-\theta)a+\theta b}{a}\,,
\end{align}
then
\begin{align}
|f(\theta)| \leq b+a-b\log b-b\log a\,.
\end{align}
\end{lemma}
\begin{IEEEproof}
It is easy to check that $f(1)=b\log\frac{b}{a}$, $f(0^{+})=b-a$ and
\begin{align} \label{eq:fd}
f'(\theta)=\frac{b-a}{\theta}-\frac{a}{\theta^2}\log\left(1-\theta+\frac{b}{a}\theta\right).
\end{align}
Define $g(\theta)=\theta(b-a)-a\log\left(1-\theta+\frac{b}{a}\theta\right)$ for $\theta\in(0,1]$, then we have
\begin{align}
g'(\theta)=\frac{\theta(b-a)^2}{(1-\theta)a+\theta b}\geq 0.
\end{align}
Since $g(0^{+})=0$, $g(\theta)\geq0$ for all $\theta\in(0,1]$. According to~\eqref{eq:fd}, we have $f'(\theta)=\frac{g(\theta)}{\theta^2}\geq0$. It follows that for all $\theta\in(0,1]$,
\begin{align}
b-a=f(0^{+})\leq f(\theta) \leq f(1)=b\log\frac{b}{a},
\end{align}
and hence
\begin{align}
|f(\theta)| &\leq\max\left\{|b-a|,\left|b\log\frac{b}{a}\right|\right\} \\
&\leq b+a-b\log b-b\log a.
\end{align}
Lemma~\ref{lm:ubd} is thus established.
\end{IEEEproof}

Applying Lemma~\ref{lm:ubd} with $a=p_Y(y;\mu_0)$ and $b=p_Y(y;\mu)$, we have
\begin{align}
&\left|\frac{1}{\theta}p_Y(y;\mu_\theta)\log\frac{p_Y(y;\mu_\theta)}{p_Y(y;\mu_0)}\right| \leq p_Y(y;\mu)+p_Y(y;\mu_0) \nonumber \\
&\quad \quad -p_Y(y;\mu)\log p_Y(y;\mu)-p_Y(y;\mu)\log p_Y(y;\mu_0)
\end{align}
where the right hand side is an integrable function of $y$ by the result that $-\int_{-\infty}^\infty p_Y(y;\mu_2)\log p_Y(y;\mu_1)\diff y<\infty$ for any $\mu_1,\mu_2\in\Lambda(q)$. In fact, as in the proof of Lemma~\ref{lm:holo} (see~\eqref{eq:logp_ub}), there exist $a,b\in\reals$ such that
$|\log p_Y(y;\mu_1)|\leq \frac{1}{2}y^2+ay+b$. Therefore,
\begin{align}
\int_{-\infty}^{\infty} |p_Y(y;\mu_2)\log p_Y(y;\mu_1)|\diff y
&\leq \int_{-\infty}^{\infty}
p_Y(y;\mu_2)\left(\frac{1}{2}y^2+ay+b\right)\diff y\\
&=\frac{1}{2}\expsub{\mu_2}{X^2}+a\expsub{\mu_2}{X}+b+\frac{1}{2}\\
&<\infty
\end{align}
due to the assumption that $\mu_2\in\Lambda(q)$.

Therefore, the dominated convergence theorem provides that
%\marginpar{\scriptsize  \parbox[t]{0.6in}{to be deleted upon acceptance}}
\begin{align}
\lim_{\theta\rightarrow0^{+}}\frac{1}{\theta}\int_{-\infty}^{\infty} p_Y(y;\mu_\theta)\log\frac{p_Y(y;\mu_\theta)}{p_Y(y;\mu_0)}\diff y %\nonumber \\
&= \int_{-\infty}^{\infty} \lim_{\theta\rightarrow0^{+}}\frac{1}{\theta} p_Y(y;\mu_\theta)\log\frac{p_Y(y;\mu_\theta)}{p_Y(y;\mu_0)}\diff y \\
&= \int_{-\infty}^{\infty} \left(p_Y(y;\mu)-p_Y(y;\mu_0)\right)\diff y \\
&= 0.
\end{align}
Lemma~\ref{lm:wd} is thus proved.
\end{IEEEproof}

We establish the following sufficient and necessary condition for the optimal input distribution.
\begin{lemma} \label{lm:N&S 1} Let
  \begin{align} \label{eq:fx}
    f_\lambda(x;\mu)=d(x;\mu)-I(\mu)-\lambda(x^2-\gamma).
  \end{align}
  Then $\mu_0 \in \Lambda(\gamma,q)$ achieves the capacity if and only if
  there exists $\lambda \geq 0$ such that $\lambda
  \expsub{\mu_0}{X^2-\gamma}=0$ and $\expsub{\mu}{f_\lambda(X;\mu_0)}\leq0$
  for all $\mu\in\Lambda(q)$.
\end{lemma}
\begin{IEEEproof}
Define the Lagrangian
\begin{align} \label{eq:6}
J(\mu)=I(\mu)-\lambda\expsub{\mu}{X^2-\gamma}
\end{align}
where $\lambda$ is the Lagrange multiplier. Since $\Lambda(q)$ is a convex set and $I(\mu)<\infty$ on $\Lambda(q)$, $\mu_0$ is capacity-achieving if and only if there exists $\lambda\geq0$ such that the following conditions hold~\cite{Luenberger69}:
\begin{enumerate}
  \item[(i)] $\lambda \expsub{\mu_0}{X^2-\gamma}=0$;
  \item[(ii)] for all $\mu \in \Lambda(q)$, $J(\mu_0)\geq J(\mu)$.
\end{enumerate}
Due to concavity of $I(\mu)$, $J(\mu)$ is also concave. Condition~(ii) is then equivalent to that the weak derivative $J'_{\mu_0}(\mu)\leq0$ for all $\mu \in \Lambda(q)$.

By Lemma~\ref{lm:wd}, the linearity of $\expsub{\mu}{X^2-\gamma}$ with
respect to (w.r.t.) $\mu$ and Condition~(i), $J'_{\mu_0}(\mu)$ can be
easily calculated as
\begin{align}
  J'_{\mu_0}(\mu)=\expsub{\mu}{f_\lambda(X;\mu_0)}.
\end{align}
Therefore, Condition~(ii) is equivalent to
$\expsub{\mu}{f_\lambda(X;\mu_0)}\leq 0$ for all
$\mu\in\Lambda(q)$. %Together with Condition~(i),
Thus Lemma~\ref{lm:N&S 1} follows.
\end{IEEEproof}

We call $x \in \reals$ a point of increase of a measure $\mu$ if
$\mu(O)>0$ for every open subset $O$ of $\reals$ containing $x$. Let
$S_{\mu}$ be the set of points of increase of $\mu$. Based on
Lemma~\ref{lm:N&S 1}, we derive another sufficient and necessary
condition for the optimal input distribution, which will be used to
prove Property~(c) of Theorem~\ref{thm:Main} in
Section~\ref{sec:propc}.

\begin{lemma} \label{lm:N&S 2} Let
  \begin{align} \label{eq:gx}
    g_\lambda(x;\mu)=qf_\lambda(0;\mu)+(1-q)f_\lambda(x;\mu).
  \end{align}
  Then $\mu_0 \in \Lambda(\gamma,q)$ achieves the capacity if and only if
  there exists $\lambda \geq 0$ such that for every $x\in\reals$,
  \begin{align}\label{eq:iff}
    g_\lambda(x;\mu_0)\leq0 \,.
  \end{align}
  Furthermore, $g_\lambda(x;\mu_0)=0$ for every $x \in
  S_{\mu_0}\backslash\{0\}$.
\end{lemma}

\begin{IEEEproof}
  The necessity part is shown as follows. Suppose $\mu_0$ achieves the
  capacity, then by Lemma~\ref{lm:N&S 1}, there exists $\lambda \geq
  0$ such that $\lambda \expsub{\mu_0}{X^2-\gamma}=0$ and
  $\expsub{\mu}{f_\lambda(X;\mu_0)}\leq0$ for all $\mu\in\Lambda(q)$. For
  any $x\in\reals\backslash\{0\}$, choose $\mu$ such that
  $\mu(\{0\})=q$ and $\mu(\{x\})=1-q$, so by the fact that
  $\mu\in\Lambda(q)$, we have
  \begin{align}\label{eq:nec}
    0\geq\expsub{\mu}{f_\lambda(X;\mu_0)}=qf_\lambda(0;\mu_0)+(1-q)f_\lambda(X;\mu_0).
  \end{align}
  Due to the continuity of $d(x;\mu_0)$ by Lemma~\ref{lm:holo},
  $f_\lambda(x;\mu_0)$ is also continuous so that~\eqref{eq:nec} holds
  for all $x \in \reals$, i.e., $g_\lambda(x;\mu_0)\leq0$ for every
  $x\in\reals$.

  To finish proving the necessity, it suffices to show that
  $g_\lambda(x;\mu_0)=0$ for all $x \in
  S_{\mu_0}\backslash\{0\}$. Evidently,
  $g_\lambda(0;\mu_0)=f_\lambda(0;\mu_0)$ and by~\eqref{eq:9} and
  $\lambda \expsub{\mu_0}{X^2-\gamma}=0$,
  \begin{align} \label{eq:intf} \int
    f_\lambda(x;\mu_0)\,\mu_0(\diff x)=0 \,.
  \end{align}
  Hence,
  \begin{align}
    \int_{\reals\backslash\{0\}} g_\lambda(x;\mu_0) \,\mu_0(\diff x) %\nonumber \\
    &= \int g_\lambda(x;\mu_0)\,\mu_0(\diff x)-g_\lambda(0;\mu_0)\mu_0(\{0\}) \\
    &\geq qf_\lambda(0;\mu_0)+(1-q)\int f_\lambda(x;\mu_0)\,\mu_0(\diff x)-qf_\lambda(0;\mu_0) \\
    &=0. \label{eq:int0}
  \end{align}
  Since $g_\lambda(x;\mu_0)\leq0$ for every $x\in\reals$,
  \eqref{eq:int0} implies that on $\reals\backslash\{0\}$,
  $g_\lambda(x;\mu_0)=0$ $\mu_0$-almost surely, so that
  $g_\lambda(x;\mu_0)=0$ for all $x \in S_{\mu_0}\backslash\{0\}$
  follows immediately.

  The sufficiency part of Lemma~\ref{lm:N&S 2} is established as
  follows. Suppose $g_\lambda(x;\mu_0)\leq0$ for every
  $x\in\reals$. By integrating $g_\lambda(x;\mu_0)$ w.r.t. $\mu_0$, we
  have
  \begin{align}
    qg_\lambda(0;\mu_0) &\geq\int g_\lambda(x;\mu_0)\,\mu_0(\diff x) \\
    &= qg_\lambda(0;\mu_0)-(1-q)\lambda\expsub{\mu_0}{X^2-\gamma} \label{eq:eq} \\
    &\geq qg_\lambda(0;\mu_0) \label{eq:ineq}
  \end{align}
  where~\eqref{eq:eq} is due to~\eqref{eq:9} and
  $g_\lambda(0;\mu_0)=f_\lambda(0;\mu_0)$, and~\eqref{eq:ineq} follows
  from $\expsub{\mu_0}{X^2}\leq \gamma$ since $\mu_0\in\Lambda(\gamma,q)$. Hence,
  $\lambda\expsub{\mu_0}{X^2-\gamma}=0$ due to the fact that
  $q<1$. Furthermore, for any $\mu\in\Lambda(q)$, by integrating
  $g_\lambda(x;\mu_0)$ w.r.t. $\mu$, we have
  \begin{align}
    qg_\lambda(0;\mu_0) &\geq\int g_\lambda(x;\mu_0) \,\mu(\diff x) \\
    &= qf_\lambda(0;\mu_0)+(1-q)\expsub{\mu}{f_\lambda(X;\mu_0)}.
  \end{align}
  Because $g_\lambda(0;\mu_0)=f_\lambda(0;\mu_0)$, we have
  $\expsub{\mu}{f_\lambda(X;\mu_0)}\leq0$. Together with
  $\lambda\expsub{\mu_0}{X^2-\gamma}=0$ and Lemma~\ref{lm:N&S 1}, this
  implies that $\mu_0$ must be capacity-achieving.
\end{IEEEproof}

\subsection{Discreteness of $\mu_0$}
\label{sec:propc}

With Lemma~\ref{lm:N&S 2} established, we now prove Property~(c) in
Theorem~\ref{thm:Main}.

Let $\lambda \geq 0$ satisfy condition~\eqref{eq:iff} and $d(z;\mu)$
be defined in~\eqref{eq:hw}. We extend functions $f_\lambda(x;\mu)$ in
Lemma~\ref{lm:N&S 1} and $g_\lambda(x;\mu)$ in Lemma~\ref{lm:N&S 2} to
be defined on the whole complex plane $\complex$ as~\eqref{eq:fx}
and~\eqref{eq:gx}, respectively, with $x$ replaced by $z\in\complex$.
\begin{comment}
  \begin{align}
    f_\lambda(z;\mu) &= d(z;\mu)-I(\mu)-\lambda(z^2-\gamma) \\
    g_\lambda(z;\mu) &= qf_\lambda(0;\mu)+(1-q)f_\lambda(z;\mu)
  \end{align}
  where $d(z;\mu)$ is defined in~\eqref{eq:hw} and $\lambda \geq 0$
  satisfies the condition~\eqref{eq:iff}.
\end{comment}
By Lemma~\ref{lm:holo}, $d(z;\mu)$ is a holomorphic function of $z$ on
$\complex$, hence so is $g_\lambda(z;\mu)$. According to
Lemma~\ref{lm:N&S 2}, each element in the set
$S_{\mu_0}\backslash\{0\}$ is a zero of the function
$g_\lambda(z;\mu_0)$.

Next we show that for any bounded interval $L$ of $\reals$, $S_{\mu_0}
\bigcap L$ is a finite set. Suppose, to the contrary, $S_{\mu_0}
\bigcap L$ is infinite, then it has a limit point in $\reals$ by the
Bolzano-Weierstrass Theorem~\cite{Lang99} and hence,
$g_\lambda(z;\mu_0)=0$ on the whole complex plane $\complex$ by the
Identity Theorem~\cite{Rudin86}. Then,
by~\eqref{eq:dxmu},~\eqref{eq:fx} and~\eqref{eq:gx}, for every
$x\in\reals$,
\begin{align} \label{eq:conv} \int_{-\infty}^{\infty}
  \phi(y-x)r(y)\diff y=0
\end{align}
where
\begin{align}
  r(y)=\log p_Y(y;\mu_0)+\lambda y^2+c
\end{align}
and $c=\frac{1}{2}\log(2\pi
e)+I(\mu_0)-\frac{q}{1-q}d(0)-\lambda(\gamma+1)$ is a constant.

As in the proof of Lemma~\ref{lm:holo}, there exist $a,b \in \reals$ such that $|\log p_Y(y;\mu_0)|\leq \frac{1}{2}y^2+ay+b$. As a result, there exist some $\alpha,\beta>0$ such that $|r(y)| \leq \alpha y^2+\beta$. Since the convolution of $r(y)$ and the Gaussian density is equal to the zero function by~\eqref{eq:conv}, $r(y)$ must be the zero function according to~\cite[Corollary~$9$]{ChaHra05IT}. This requires the capacity-achieving output distribution $p_Y(y;\mu_0)$ be Gaussian, which cannot be true unless $X$ is Gaussian, which contradicts the assumption that $X$ has a probability mass at 0. Therefore, $S_{\mu_0}\bigcap L$ must be a finite set for any bounded interval $L$, which further implies that $S_{\mu_0}$ is at most countable.

Finally, we show that $S_{\mu_0}$ is countably infinite. Suppose, to
the contrary, $S_{\mu_0}=\{x_i\}_{i=1}^N$ is a finite set with
$\mu_0(\{x_i\})=p_i$ and $|x_i|\leq B_1$ for all $i=1,2,\dots,N$. For
any $y>B_1$,
\begin{align}
  p_Y(y;\mu_0) = \sum_{i=1}^N p_i\phi (y-x_i)
  \leq e^{-\frac{(y-B_1)^2}{2}} \ .
\end{align}
For any $\epsilon>0$, choose $B_2>0$ such that
$\int_{-B_2}^{B_2}\phi(x)\diff
x>1-\epsilon$. By~\eqref{eq:dxmu},~\eqref{eq:fx},~\eqref{eq:gx}
and~\eqref{eq:iff}, for any $x>B_1+B_2$, we have
\begin{align}\label{eq:inf_lb}
  0 &\geq -\int_{-\infty}^{\infty}
  \phi(y-x) \log p_Y(y;\mu_0)\diff y -\lambda x^2-(c+\lambda) \\
  &\geq \int_{x-B_2}^{x+B_2}
  \phi(y-x) \frac{1}{2}(y-B_1)^2\diff y -\lambda x^2-(c+\lambda) \\
  &= \int_{B_2}^{B_2} \phi(t) \frac{1}{2}(x-B_1+t)^2\diff t -\lambda x^2-(c+\lambda) \\
  &\geq \frac{1}{2}(x-B_1)^2(1-\epsilon)-\lambda x^2-(c+\lambda).
\end{align}
For~\eqref{eq:inf_lb} to hold for large $x$, $\lambda$ must satisfy
$\lambda \geq \frac{1}{2}$.

To finish the proof, it suffices to show that $\lambda<\frac{1}{2}$
for any $\gamma>0$, so that contradiction arises, which implies that
$S_{\mu_0}$ must be countably infinite. For fixed $q \in (0,1)$,
denote the Lagrange multiplier in~\eqref{eq:iff} as
$\lambda(\gamma)$. Denote $C_G(\gamma)=\frac{1}{2}\log(1+\gamma)$, which is the
channel capacity of a Gaussian channel with the average power
constraint only. By the envelope theorem~\cite{Luenberger69},
$\lambda(\gamma)$ is the derivative of $C(\gamma,q)$ w.r.t. $\gamma$. Since
$C(0,q)=C_G(0)=0$ and the derivative of $C_G(\gamma)$ at $\gamma=0$ is
$\frac{1}{2}$, we have $\lambda(0) \leq \frac{1}{2}$, otherwise we
could find a small enough $\gamma$ such that $C(\gamma,q)$ would exceed $C_G(\gamma)$
which is obviously impossible. Next we show that $C(\gamma,q)$ is strictly
concave for $\gamma\geq 0$. Suppose $\mu_1$ and $\mu_2$ are the
capacity-achieving input distributions of~\eqref{eq:Capacity1} for
different power constraints $\gamma_1$ and $\gamma_2$, respectively. Due to
Property~(b) in Theorem~\ref{thm:Main}, $\mu_1$ and $\mu_2$ must be
different.  Define $\mu_\theta=\theta\mu_1+(1-\theta)\mu_2$ for
$\theta \in (0,1)$. It is easy to see that $\mu_\theta$ satisfies that
the duty cycle is no greater than $1-q$ and the average input power is
no greater than $\theta \gamma_1+(1-\theta)\gamma_2$. Now we have
\begin{align}
  C(\theta \gamma_1+(1-\theta)\gamma_2,q) &\geq I(\mu_\theta) \\
  &> \theta I(\mu_1)+(1-\theta)I(\mu_2) \label{eq:SConc} \\
  &= \theta C(\gamma_1,q)+(1-\theta)C(\gamma_2,q),
\end{align}
where~\eqref{eq:SConc} is due to the strict concavity of
$I(\mu)$. Therefore, the strict concavity of $C(\gamma,q)$ for $\gamma\geq 0$
follows, which implies that $\lambda(\gamma)<\lambda(0)=\frac{1}{2}$ for
all $\gamma>0$.

\section{Proof of Theorem \ref{thm:lowerbound} (the Case of Realistic
  Duty Cycle Constraint)}
\label{sec:lower-bound-capacity}

\subsection{Stationarity of the Capacity-achieving Input Distribution }
\label{sec:cons-about-stat}
We first establish the fact that a stationary distribution achieves
the capacity of the AWGN channel with the realistic duty cycle
constraint and power constraint.
\begin{comment}
 For the AWGN
channel (\ref{eq:Y=X+N}), we have
\begin{align}
  \label{eq:5n}
  p_{Y^n|X^n}(y^n|x^n)=\prod_{i=1}^np_{Y|X}(y_i|x_i)=\prod_{i=1}^n\phi(y_i-x_i).
\end{align}
\end{comment}

\begin{proposition}\label{prop:stationary}
    A stationary distribution\footnote{The stationarity of distribution $\nu$ on
    $X^n$ satisfies
    \[ \nu_{X_s,\cdots,X_t}=\nu_{X_{s+k},\cdots,X_{t+k}} \]
    for any index $s,t,k$ satisfied
    \[ 1\le s\le t\le n \quad 1\le s+k \le t+k \le n\]
  } achieves
  \begin{align}
    \max_{\mu\in \Lambda^n(\gamma,q,c)} I(X^n;Y^n).
    \label{eq:27}
  \end{align}
\end{proposition}
\begin{IEEEproof}
  Let $T_k(\cdot)$ as a $k$-cyclic-shift operator on
  $\mu\in\Lambda^n(\gamma,q,c)$, defined as
  \begin{align}
    \label{eq:26}
    T_k(\mu)=\mu_{X_{k+1},\cdots,X_n, X_1,\cdots X_k}
  \end{align}
  where $k=1,\cdots,n-1$, and specifically $T_0(\mu)=\mu$.  For
  any distribution $\mu$ in $\Lambda^n(\gamma,q,c)$, a distribution
  $\nu$ on $X^n$ can be defined as
  \begin{align}
    \label{eq:21}
    \nu=\frac{1}{n}\;\sum_{k=0}^{n-1}T_k(\mu).
  \end{align}
  According the concavity of the mutual information $I(\cdot)$,
  \begin{align}
    I(\nu)&=I\left(\frac{1}{n}\;\sum_{k=0}^{n-1}T_k(\mu)\right) \\
    &\ge \frac{1}{n}\;\sum_{k=0}^{n-1} I(T_k(\mu)) \\
    &=I(\mu)
  \end{align}
  where $I(T_k(\mu))=I(\mu)$ since the AWGN channel (\ref{eq:Y=X+N})
  is a memoryless and time-invariant.
  Obviously $\nu$ is a stationary distribution and satisfied the duty
  cycle constraint and power constraint, i.e., $\nu\in\Lambda^n(\gamma,q,c)$,
  hence Proposition \ref{prop:stationary} established.
\end{IEEEproof}

According to Proposition \ref{prop:stationary}, for any $n$,
$I(X^n;Y^n)$ is maximized by a stationary distribution. Therefore with
$n$ converges to infinity, the capacity in (\ref{eq:11}) is achieved
by a stationary input distribution.

\subsection{The Input-output  Mutual Information}

% The capacity of the AWGN channel (\ref{eq:Y=X+N}) with
% duty cycle constraint $(q,c)$ and SNR constraint $P$ is
% \begin{align}
%   \label{eq:11}
%   C=\max_{\mu\in \Lambda(P,q,c)} \lim_{n\to\infty} \frac{1}{n} I(X^n;Y^n).
% \end{align}

\begin{proposition} \label{prop:1} Let the input follows a stationary
  distribution $\mu\in\Lambda(\gamma,q,c)$. The limit of the input-output
  mutual information per symbol as a function of $\mu$ can be
  expressed as
  \begin{align}
    \label{eq:17n}
%    I(\mu)=\frac{1}{2}\log(1+P)-D(P_Y\|P_{Y_g})-h(Y_1)+h({\mathscr{Y}}).
    I(\mu)=I(X;Y)-h(Y)+h({\mathscr{Y}})
  \end{align}
  where $I(X;Y)$ is the mutual information of the AWGN channel between
  the input $X$, which follows distribution $\mu_{X_1}$ and the
  corresponding output $Y$, $h(Y)$ is the differential entropy of $Y$
  and $h({\mathscr{Y}})$ is the differential entropy rate of output
  process $\{Y_i\}$.
  \begin{comment}
    $Y_g$ is Gaussian random variables
    with $Y_{g}\sim P_{Y_g}={\mathcal{N}}(0,\gamma+1)$,
    $P_Y$ is the time-homogeneous marginal probability distribution of
    $Y_i$, $h(Y_1)$ is the differential entropy of $Y_1$ and
    $h({\mathscr{Y}})$ is the differential entropy rate of $\{Y_i\}$.
  \end{comment}
\end{proposition}
\begin{IEEEproof}
  The mutual information  between $X^n$ and $Y^n$ can
  be expressed using relative entropies
  \begin{align}
    I(X^n;Y^n) % \nonumber \\
    &=D(P_{Y^n|X^n}\|P_{Y^n}|P_{X^n}) \\
    &=D(P_{Y^n|X^n}\|P_{Y_1}\times\cdots\times
    P_{Y_n}|P_{X^n})-D(P_{Y^n}\|P_{Y_1\times\cdots\times
      P_{Y_n}})     \label{eq:13n} \\
    &=\sum_{k=1}^nD(P_{Y_k|X_k}|P_{Y_k}|P_{X^n}) -{\mathsf{E}}\left\{
      \log{P_{Y^n}(Y^n)}-\sum_{i=1}^n\log{P_{Y_i}(Y_i)}
    \right\} \\
    &= nI(X;Y)-nh(Y)+h(Y^n).
  \end{align}
  Then
  \begin{align}
      I(\mu) &= \lim_{n\to\infty}\frac{1}{n}I(X^n;Y^n) \\
      &=I(X;Y)-h(Y)+\lim_{n\to\infty}\frac{1}{n}h(Y^n) \\
%      &=I(X;Y)-h(Y)+\lim_{n\to\infty}h(Y_1|Y_2^n)  \\
      &=I(X;Y)-h(Y)+h({\mathscr{Y}}).    \label{eq:33n}
  \end{align}
  Proposition \ref{prop:1} is established.
\end{IEEEproof}

When the input is an i.i.d. random process, the output process is also
i.i.d., $h(Y)=h({\mathscr{Y}})$.  This implies the following
corollary.
\begin{corollary}
  Among all i.i.d. distributions, the one that maximizes the mutual
  information under duty cycle constraint $(q,c)$ and
  average power constraint $\gamma$ can be solved from the following
  optimization:
  \begin{align}
    \label{eq:16n}
    \begin{split}
      &\underset{P_X}{\mathrm{maximize}}\quad I(X;Y) \\
      &\begin{array}{ll} {\mbox{subject to}}& P_X(0)-2cP_{X}(0)(1-P_{X}(0))\le q,\\
        &{\mathsf{E}}\{X^2\}\le \gamma.
      \end{array}
    \end{split}
  \end{align}
\end{corollary}
In the special case of no transition cost, i.e., $c=0$, the result of
(\ref{eq:16n}) is equal to that of (\ref{eq:Capacity1}).

\subsection{Proof of Theorem \ref{thm:lowerbound}}
\label{sec:proof-theorem-lowerbound}

The mutual information expressed by (\ref{eq:17n}) is hard to optimize,
even if the input is restricted to Markov processes. To simply the
matter, we introduce a lower bound of $I(\mu)$, which is given by
$L(\mu)$ in (\ref{eq:21n}).
%\begin{definition}
  % We define
  % \begin{align}
  %   \label{eq:21n}
  %   L(\mu)=I(X;Y)-h(X)+h({\mathscr{X}})
  % \end{align}
%\end{definition}

\paragraph*{Property (a)}
\label{sec:property-lower-bound}

% \begin{proposition}\label{prop:2}
%   For any $\mu \in \Theta(\gamma,q,c)$,
%   \begin{align}
%     \label{eq:22n}
%     I(\mu) \ge I(X;Y)-h(X)+h({\mathscr{X}}).
%   \end{align}
%   For convenience, let $L(\mu)$ denote the lower bound:
%     \begin{align}
%     \label{eq:21n}
%     L(\mu)=I(X;Y)-h(X)+h({\mathscr{X}})
%   \end{align}
% \end{proposition}
%\begin{IEEEproof}
  Using the fact that processing reduce relative entropy and $\mu$ is
  specified as a stationary probability distribution, we have
  \begin{align}
    \frac{1}{n}D(P_{Y^n}\|P_{Y_{1}}\!\times\!
    P_{Y_{2}}\!\times\!\cdots\!\times\!  P_{Y_{n}}) %\nonumber \\
    &\le \frac{1}{n}D(P_{X^n}\|P_{X_{1}}\!\times\!
    P_{X_{2}}\!\times\!\cdots\!\times\!  P_{X_{n}}) \\
    &=\frac{1}{n}\sum_{k=1}^{n-1}D(P_{X_k|X_{k+1}^n}\|P_{X_k}|P_{X_{k+1}^n})    \\
    &=\frac{1}{n}\sum_{k=2}^{n}I(X_1;X_2^k).\label{eq:22}
%    &=I(X_1;X_2^\infty) \label{eq:6n}
    % &=\;\;h(X)-h({\mathscr{X}}).
  \end{align}
  Therefore
  \begin{align}
    \label{eq:6n}
    \lim_{n\to\infty}\frac{1}{n}D(P_{X^n}\|P_{X_{1}}\!\times\!
    P_{X_{2}}\!\times\!\cdots\!\times\!  P_{X_{n}})=I(X_1;X_2^\infty)
  \end{align}
  using the fact that the Ces\'aro mean of sequence $I(X_1,X^k_2)$ is
  $I(X_1;X_2^\infty)$. Applying (\ref{eq:17n}), (\ref{eq:13n}) and
  (\ref{eq:6n}),
  \begin{align}
    \label{eq:23}
    L(\mu)=I(X;Y)-I(X_1;X_2^\infty)\le I(\mu) \le C(\gamma,q,c).
  \end{align}
  Thus Property (a) is established.
%\end{IEEEproof}

% \begin{theorem}\label{thm:2}
%   The maximum of $L(\mu)=I(X;Y)-h(x)+h(\mathscr{X})$ with
%   $\mu\in\Theta(\gamma,q,c)$ is achieved by a discrete first-order Markov input
%   process. Moreover this Markov process satisfies the following
%   property: For every $i$, let $B_i=1_{\{X_i\ne0\}}$ to be the
%   indicator of $X_i$, then conditioned on $B_i$ and $B_{i+1}$, the
%   variables $X_i$ and $X_{i+1}$ are independent and
%   \begin{align}
%     \label{eq:1}
%     L(\mu)=I(X;Y)-I(B_1;B_2)
%   \end{align}  .
% \end{theorem}

% \begin{IEEEproof}

  \paragraph*{Property (b)}

  For any $\mu\in\Lambda(\gamma,q,c)$, which is not Markov in general, its
  first-order Markov approximation $\nu$ is defined by
  \begin{align}
    \label{eq:23n}
    \nu_{X_1,\cdots,X_n}=\mu_{X_1}\mu_{X_{2}|X_{1}}\mu_{X_{3}|X_{2}}\cdots \mu_{X_n|X_{n-1}}.
  \end{align}
  Evidently, $\nu$ and $\mu$ have identical marginal distributions:
  $\nu_{X_i}=\mu_{X_i}$, and also identical joint distributions of any
  consecutive pairs: $\nu_{X_i,X_{i+1}}=\mu_{X_i,X_{i+1}}$. Therefore
  \begin{align}
      \nu_{X_i}(\{0\})=\mu_{X_i}(\{0\})
  \end{align}
and
  \begin{align}
      \nu_{X_i,X_{i+1}}(\{x_i=0,x_{i+1}\ne0\})
      =\mu_{X_i,X_{i+1}}(\{x_i=0,x_{i+1}\ne0\}.
  \end{align}
  Since $\mu\in\Lambda(\gamma,q,c)$, we have $\nu\in \Lambda(\gamma,q,c)$. Let
  $\{X_i\}$ follow distribution $\mu$ and $\{Z_i\}$ follow
  distribution $\nu$. Then
  \begin{align}
      I(Z_1;Z_2^\infty)&=I(Z_1;Z_2)+I(Z_1;Z_3^\infty|Z_2) \\
      &=I(Z_1;Z_2) \\
      &=I(X_1;X_2) \\
      &\le I(X_1;X_2^\infty) \label{eq:27n}
      % h({\mathscr{X}})&=\lim_{n\to\infty}\frac{1}{n}\sum_{i=1}^nh(X_i|X_1,X_2\cdots,X_{i-1})\\
      % \le&\lim_{n\to\infty}\frac{1}{n}\sum_{i=1}^{n}h(X_i|X_{i-1})\\
      % &=\lim_{n\to\infty}\frac{1}{n}\sum_{i=1}^{n}h(Z_i|Z_{i-1})\\
      % &=h({\mathscr{Z}})
  \end{align}
  where equality holds if and only if $\{X_i\}$ is a first-order
  Markov process.  By (\ref{eq:21n}) and (\ref{eq:27n}), $L(\nu)\ge
  L(\mu)$. So for any $\mu$ which maximizes $L(\mu)$, $\nu$ can be
  generated from $\mu$ by (\ref{eq:23n}) with $L(\nu)\ge L(\mu)$.
  $L(\mu)$ must be maximized by a first-order Markov process.

  \paragraph*{Property (c)}
  Suppose $\nu$ is a stationary fist-order Markov process,
  sufficiently denote as $\nu=\{\mathcal{X}, P_{X_2|X_1}\}$, where $\mathcal{X}$ is
  the state space of $\nu$ and $P_{X_2|X_1}$ is the transition
  probability distribution.  Define a new first-order Markov process
  $\bar{\nu}$ from $\nu$ as follows.
  \begin{definition}\label{def:2}
    Let $\bar{\nu}$, defined on the same state space $\mathcal{X}$ as
    $\nu$, be a first-order Markov process denoted by $(\mathcal{X},
    P_{Z_2|Z_1})$, where
    \begin{align}
      \label{eq:28n}
      P_{Z_2|Z_1}(z_2|z_1)=
      \begin{cases}
        \alpha & z_1=0\; z_2=0, \\
        1-\beta & z_1\ne 0\; z_2=0, \\
        \dfrac{1-\alpha}{\eta}P_X(z_2) & z_1=0\; z_2\ne0,\\
        \dfrac{\beta}{\eta} P_X(z_2) & z_1\ne0\; z_2\ne0,
      \end{cases}
    \end{align}
    where
    \begin{align}
      \label{eq:30}
      S_1=\mathcal{X}\setminus \{0\}
    \end{align}
    and
    \begin{align}
      \alpha&=P_{X_2|X_1}(0|0) \\
      \beta&=P(X_2\in S_1|X_1\in S_1) \\
      \eta&=P(X\in S_1)        .
    \end{align}
  \end{definition}

  The process $\bar{\nu}$ is described by $(\mathcal{X}, \alpha, \beta,
  P_X)$. It is easy to prove that the stationary distribution $P_Z$ of
  $\bar{\nu}$ is equal to $P_X$ of $\nu$,
  $\bar{\nu}\in\Lambda(\gamma,q,c)$. Moreover, $\bar{\nu}$ satisfies the
  same power and duty cycle constraint $\nu$ satisfies, i.e.,
  $\bar{\nu}\in\Lambda(\gamma,q,c)$. Furthermore let $B_i=1_{\{X_i\ne0\}}$, then
  \begin{align}
    P_{B_2|B_1}(0|0)&=\alpha  \\ %&P_{B_2|B_1}(1|0)&=1-\alpha \nonumber\\
    % P_{B_2|B_1}(0|1)&=1-\beta &
    P_{B_2|B_1}(1|1)&=\beta  . %\nonumber
  \end{align}
  Let $b_i=1_{\{z_i\ne0\}}$. Since
  \begin{align}\label{eq:29}
    P_{Z_2|Z_1}(z_2|z_1)=P_{B_2|B_1}(b_1|b_2)\frac{P_{X}(z_2)}{P_{B_2}(b_2)},
  \end{align}
  $Z_i$ and $Z_{i+1}$ are independent given $B_i=1_{\{Z_i\ne0\}}$ and
  $B_{i+1}=1_{\{Z_{i+1}\ne0\}}$.

  Based on (\ref{eq:28n}) to (\ref{eq:29}), it is easy to see that
  \begin{align}
    % h({\mathscr{Z}})&=h(X)+(1-\gamma)H_2(\alpha)+\gamma
    % H_2(\beta)-H_2(\gamma)\nonumber\\
    % &=h(X)-I(B_1;B_2)\nonumber \\
    % \ge& h(X)-I(X_1;X_2) \nonumber \\
    % &=h({\mathscr{X}})
    I(Z_1;Z_2)&=\expect{\log \frac{P_{Z_2|Z_1}(Z_2|Z_1)}{P_{Z_2}(Z_2)}} \\
    &=\expect{\log\frac{P_{B_2|B_1}(B_2|B_1)}{P_{B_2}(B_2)}} \\
    &=I(B_1;B_2) \\
    &\le I(X_1;X_2).
    \label{eq:32n}
  \end{align}
  The inequality in (\ref{eq:32n})
  follows since $X_1\to X_2\to B_2$ forms a Markov chain then
  $I(X_1;B_2)\le I(X_1;X_2)$ \cite{Cover2006} and $B_2\to X_1\to B_1$
  also forms a Markov chain then $I(B_2;B_1)\le I(B_2;X_1)$.

  The discreteness of the optimized input distribution is proved in
  the following. According to Properties (b) and (c), lower bound
  $L(\cdot)$ is maximized by a first-order Markov process, the
  transition probability distribution of which $P_{X_2|X_1}$ can be
  expressed as
  \begin{align}
    \label{eq:5}
    P_{X_2|X_1}(x_2|x_1)=P_{B_2|B_1}(b_2|b_1)\frac{P_{X}(x_2)}{P_{B_2}(b_2)}
  \end{align}
  where $b_i=1_{\{x_i\ne0\}}$ $P_X=\mu_{X}$ and
  $P_{X_2|X_1}=\mu_{X_2|X_1}$. Then the maximum of $L(\mu)$ can be
  achieved by the follow optimization
  \begin{align}
    \underset{q_0}{\mathrm{maximize}}\hspace{1em}&\quad
    I_X(q_0)-I_B(q_0) \label{eq:25} \\
    \mbox{subject to \hspace{1em}}
     &I_X(q_0)=\underset{P_X}{\mathrm{maximize}}\quad I(X;Y) \label{eq:32} \\
     & I_B(q_0)=\underset{P(B_2|B_1)}{\mathrm{minimize}}\quad
     I(B_1;B_2)\\
     &P_X(0)=P_{B_1}(0)=P_{B_2}(0)=q_0 \\
     &q_0-2cq_0P_{B_2|B_1}(1|0)\ge q. %\\
%     &q_0P_{B_2|B_1}(0|0)+(1-q_0)P_{B_2|B_1}(0|1)=q_0 \label{eq:20}
  \end{align}
  %where (\ref{eq:20}) guarantees that $P_{B_1}=P_{B_2}$.
  Since given
  any $q_0\ge q>0$, $I_X(q_0)-I_B(q_0)$ can be maximized by the
  maximum of $I_X(q_0)$ and the minimum of $I_B(q_0)$ respectively,
  the maximization of (\ref{eq:25}) must be achieved by $P_X$, which
  maximizes $I(X;Y)$ for given $q_0$. Therefore given $q_0$, the
  maximization in (\ref{eq:32}) is similar to the problem in Theorem
  \ref{thm:Main}. The difference to Theorem \ref{thm:Main} is that in
  (\ref{eq:32}) the distribution $P_X$ satisfies $P_X(0)=q_0\ge q$,
  however in Theorem \ref{thm:Main} the distribution $P_X$ satisfies
  $P_X(0)\ge q$. Define
  \begin{align}
  \Lambda_0(\gamma,q_0)&=\big\{\mu:\mu(\{0\})=q_0, \,
  \expsub{\mu}{X^2}\leq \gamma\}
  \end{align}
  where $\mu$ is the marginal input distribution of the first-order Markov process. We can establish the following lemma.
  \begin{lemma}\label{lm:compact2}
    $\Lambda_0(\gamma,q_0)$ is compact in the topological space $\mathcal{P}$.
  \end{lemma}
  \begin{IEEEproof}
  As mentioned in Lemma~\ref{lm:compact}, the topology of weak convergence on $\mathcal{P}$ is metrizable with the L\'{e}vy-Prohorov metric~\cite{Stroock93} and defined as
  \begin{align}\label{eq:L}
  L(\mu,\nu) = \inf\big\{\delta: &\mu(A)\leq\nu(A^{(\delta)})+\delta\ \text{and}\ \nonumber \\
  &\nu(A)\leq\mu(A^{(\delta)})+\delta\ \text{for all}\ A \subseteq \mathcal{B} \big\}
  \end{align}
  for any $\mu,\nu \in \mathcal{P}$, where $A^{(\delta)}$ denotes the set of all $x \in \reals$ which lie a $d$-distance less than $\delta$ from $A$.

  Similarly as in the proof of Lemma~\ref{lm:compact}, it suffices to show that $\Lambda_0(\gamma,q_0)$ is both tight and closed in $\mathcal{P}$. The tightness can be shown by the same arguments as in Lemma~\ref{lm:compact}. In the following, we prove that $\Lambda_0(\gamma,q_0)$ is closed in $\mathcal{P}$.

  Let $B_m=\left[-\frac{1}{m},\frac{1}{m}\right]$ for $m=1,2,\dots$. Let $\{\mu_n\}_{n=1}^{\infty}$ be a convergent sequence in $\Lambda_0(\gamma,q_0)$ with limit $\mu_0$. For any $m\in\mathbb{N}$, there exists an $n_m$ such that $L(\mu_n,\mu_0)<\frac{1}{m}$ for all $n>n_m$. By the definition of $L$ in~\eqref{eq:L}, we have for any $m\in\mathbb{N}$ and $n>n_m$,
  \begin{equation}\label{eq:mu0ub}
  \mu_0(\{0\}) \leq \mu_n(B_m)+\frac{1}{m}\,,
  \end{equation}
  and
  \begin{equation}\label{eq:mu0lb}
  \mu_n(\{0\}) \leq \mu_0(B_m)+\frac{1}{m}\,.
  \end{equation}
  For any $n\in\mathbb{N}\bigcup\{0\}$, we have
  \begin{equation}
  \mu_n(\{0\})=\mu_n\left(\bigcap_{m=1}^{\infty}B_m\right)=\lim_{m\rightarrow\infty}\mu_n(B_m),
  \end{equation}
  so for any $m\in\mathbb{N}$, there exists an $n_m'$ such that $\mu_n(B_m)\leq\mu_n(\{0\})+\frac{1}{m}$. Therefore, according to~\eqref{eq:mu0ub} and~\eqref{eq:mu0lb}, for all $m\in\mathbb{N}$ and $n>\max\{n_m,n_m'\}$,
  \begin{equation}
  q_0-\frac{2}{m} \leq \mu_0(\{0\}) \leq q_0+\frac{2}{m}.
  \end{equation}
  Thus we have $\mu_0(\{0\})=q_0$ by letting $m\rightarrow\infty$.

  Moreover, let $f(x)=x^2$ which is continuous and bounded below. By weak convergence~\cite[Section~$3.1$]{Stroock93}, we have
  \begin{equation}
  \expsub{\mu_0}{X^2}=\int f\diff\mu_0\leq \liminf_{n\rightarrow\infty} \int f\diff\mu_n \leq \gamma.
  \end{equation}
  Together with $\mu_0(\{0\})=q_0$, we have $\mu_0 \in \Lambda_0(\gamma,q_0)$, i.e., $\Lambda_0(\gamma,q_0)$ is closed, and the compactness of $\Lambda_0(\gamma,q_0)$ then follows.
  \end{IEEEproof}

  Now $P_X$ can be proved to be discrete by following the same
  development as in the proof of Theorem~\ref{thm:Main} with
  Lemma~\ref{lm:compact} substituted by Lemma~\ref{lm:compact2}.
  Because $P_X$ is
  the stationary distribution of the Markov process, the maximum of
  the lower bound $L(\cdot)$ is achieved by a discrete first-order
  Markov process.

  Based on Theorem~\ref{thm:lowerbound}, in order to find the lower
  bound of the capacity, we can maximize $L(\mu)$ and obtain an
  optimized discrete first-order Markov input $\mu^*=\{\mathcal{X}, \alpha,
  \beta, P_X\}$ in $\Lambda(\gamma,q,c)$. Let $\mu_0$ denote the
  capacity-achieving distribution, then
\begin{align}
  \label{eq:31n}
  I(\mu_0)\ge I({\mu^*}) \ge L({\mu^*}).
\end{align}

In Section \ref{sec:numer-comp-entr}, we develop a computationally
efficient scheme to determine $\mu^*$, which is a good approximation
of  the capacity-achieving input $\mu_0$.

\section{Numerical methods and results}
\label{sec:numer-meth-results}

\subsection{Computation of the entropy of Hidden Markov Processes}\label{sec:numer-comp-entr}

In order to numerically calculate the % lower bounder of
mutual information~\eqref{eq:17n}, it is important to
compute the differential entropy rate of a HMP generated by Markov
input through the AWGN channel.
Computing the (differential) entropy rate of HMPs is a hard problem.
% Due to the hardness in theory, it is still an open question on the
%  efficient computation of entropy rate of HMP.
Most works in this area focus on the entropy rate of the binary Markov
input through various channels. Reference \cite{Ordentlich2005}
solves a linear system for the stationary distribution of the
quantized Markov process to obtain a good approximation of the entropy
rate for the HMP output generated by binary Markov input through a
binary symmetric channel. In \cite{Luo2009IT}, the entropy rate of
HMP generated by binary-symmetric Markov input through arbitrary
memoryless channels is studied and a numerical method is presented
based on quantizing a fixed-point functional equation.  Based on these
existing studies,
% above research work,
a Monte Carlo algorithm is provided in this paper to compute the
differential entropy rate of HMPs generated from a $m$-state
Markov chain ($m\ge3$) through the AWGN channel.  We sketch the main ideas in
our algorithm for computing the differential entropy rate in this subsection.

Based on Blackwell's work\cite{Blackwell1957}, the entropy of HMPs can
be expressed as an expectation on the distribution of the conditional
distribution of $X_0$ given the past observations $Y_{-\infty}^0$.  In
order to estimate $P_{X_0|Y_{-\infty}^0}$, first define the
log-likelihood ratio:
\begin{align}
  \label{eq:7}
    L_n^{(i)}=\log\frac{P_{X_n|Y^n}(X^{(i)}|Y^n)}{P_{X_n|Y^n}(X^{(0)}|Y^n)},
    \quad i=0,1,\cdots, m-1
\end{align}
where $m$ is the number of the states of Markov Chain, $X^{(i)}\in
\mathcal{X}$ is the $i$th state and $\mathcal{X}$ is the state space of Markov
Chain. It is obviously that $L_n^{(0)}\equiv 0$. Then given
$\vect{L}_{n}=\{L_{n}^{(0)}, L_{n}^{(1)},\cdots, L_{n}^{(m-1)}\}$,
$P_{X_n|Y^n}(X_n|Y^n)$ can be calculated as
\begin{align}
  \label{eq:16}
  P_{X_{n}|Y^n}(X^{(i)}|Y^n)
  =\frac{e^{L^{(i)}_n}}{\sum_{i=0}^{m-1}e^{L^{(i)}_n}}
\end{align}
and when $n\to\infty$, (\ref{eq:16}) converges to $P_{X_0|Y^0_{-\infty}}(X^{(i)}|Y^0_{-\infty})$.

In addition, $L_{n+1}^{(i)}$ can be calculated from $\vect{L}_{n}$
 iteratively as
\begin{align}
  \label{eq:14}
  L_{n+1}^{(i)}=R^{(i)}(Y_{n+1})+F^{(i)}(\vect{L}_n)
\end{align}
where
\begin{align}
 \label{eq:13}
 R^{(i)}(Y_{n+1})&=(X^{(i)}-X^{(0)})\;Y_{n+1}-\frac{1}{2}((X^{(i)})^2-(X^{(0)})^2)\\
 F^{(i)}(\vect{L_{n}})&=\log\frac{\sum_{k=0}^{m-1}P_{X_2|X_1}(X^{(i)}|X^{(k)})\;e^{L_n^{(k)}
   } }{\sum_{k=0}^{m-1}P_{X_2|X_1}(X^{(0)}|X^{(k)})\;e^{L_n^{(k)} }}.
\end{align}
Detail deduction of (\ref{eq:14}) is shown in (\ref{eq:8})
\begin{align}
  \label{eq:8}
  L_{n+1}^{(i)}
  &=\log\frac{P_{X_{n+1}|Y_1^{n+1}}(X^{(i)}|Y_1^{n+1})}{P_{X_{n+1}|Y_1^{n+1}}(X^{(0)}|Y_1^{n+1})}\\
  &=\log\frac{\sum_{k=0}^{m-1}
      P_{Y_{n+1}|X_{n+1}}(Y_{n+1}|X^{(i)})P_{X_2|X_1}(X^{(i)}|X^{(k)})P_{X_n|Y_1^n}(X^{(k)}|Y_1^n)}{\sum_{k=0}^{m-1}
      P_{Y_{n+1}|X_{n+1}}(Y_{n+1}|X^{(0)})P_{X_2|X_1}(X^{(0)}|X^{(k)})P_{X_n|Y_1^n}(X^{(k)}|Y_1^n)}\\
    &=\log\frac{P_{Y_{n+1}|X_{n+1}}(Y_{n+1}|X^{(i)})}{P_{Y_{n+1}|X_{n+1}}(Y_{n+1}|X^{(0)})}
    +\log\frac{\sum_{k=0}^{m-1}P_{X_2|X_1}(X^{(i)}|X^{(k)})\;e^{L_n^{(k)} }
    }{\sum_{k=0}^{m-1}P_{X_2|X_1}(X^{(0)}|X^{(k)})\;e^{L_n^{(k)} }}\\
    &=R^{(i)}(Y_{n+1})+F^{(i)}(\vect{L}_n).
\end{align}

For the hidden Markov processes observed through the
AWGN channel~(\ref{eq:Y=X+N}), the entropy of HMPs can be computed as \cite{Blackwell1957}
\begin{align}
  \label{eq:15}
  \begin{split}
    h(\mathscr{Y})&=\lim_{n\to\infty}-\!\iint r(y, \vect{l}_{n})\log r(y, \vect{l}_{n}) \; dy\;dP_{\vect{L}_n}\!(\vect{l}_{n})
  \end{split}
\end{align}
where
\begin{align}
  \label{eq:17}
  \begin{split}
    &r(y,\vect{L}_{n})
    =
    \sum_{i=0}^{m-1}\phi(y-x^{(i)})\sum_{k=0}^{m-1}\frac{e^{L^{(i)}_n}}{\sum_{i=0}^{m-1}e^{L^{(i)}_n}} P_{X_2|X_1}(x^{(i)}|x^{(k)}).
  \end{split}
\end{align}

In order to compute the entropy rate of HMPs based on (\ref{eq:15}),
the key is to estimate the probability distribution of $\vect{L}_n$,
$P_{\vect{L}_n}$. In \cite{Ordentlich2005} for binary Markov input and
the binary symmetric channel, $\vect{L}_n$ is considered as a
1-dim $M$-state Markov chain by quantizing the dynamic system
expressed in (\ref{eq:14}). Then the distribution of $\vect{L}_\infty$
is the stationary distribution of the quantized Markov process and can
be computed easily through eigenvector solving method. In this paper
because the number of states of the Markov input, $m$ is larger than 2
and the HMPs is observed through the AWGN channel, directly quantizing
the dynamic system (\ref{eq:14}) will generate a quantized Markov
chain with $M^{m-1}$ states, which is very difficult to deal with when
large $M$ is selected for good estimation precision.

According to (\ref{eq:14}), since $\vect{L}_{n+1}$ is only dependent
on $\vect{L}_n$ and $Y_{n+1}$, $\{\vect{L}_n\}$ can be considered as a
Markov process. In order to compute the stationary probability
distribution $P_{\vect{L}_\infty}$, we can evolve the distribution of
$\vect{L}_n$ based on (\ref{eq:14}) from any initial distribution
$P_{\vect{L}_0}$. When $n$ is large enough, the distribution
$P_{\vect{L}_n}$  converges to $P_{\vect{L}_\infty}$. A Monte Carlo algorithm
for approximating $h(\mathscr{Y})$ is introduced as follows:
\begin{enumerate}[1)]
\item Initialize $M$ particles $\{\vect{L}_{0,1},\cdots,\vect{L}_{0,M}\}$,
  $\vect{L}_{0,k}$ can be simply sampled from the $(m\!-\!1)$-dim Uniform
  distribution with each dimension on $[-\max(X^{(i)}), \max(X^{(i)})]$.
\item for $n=0,1,2,\cdots,N$, iteratively evolve the particles
  $\{\vect{L}_{0,1},\cdots,\vect{L}_{0,M}\}$ based on (\ref{eq:14}),
  where each $y_{n+1,k}$ is sampled according to $r(y, \vect{L}_{n, k})$.
\item when $N$ is large enough,  $\{\vect{L}_{N,k}\}$ can be used to
  estimate $h(\mathscr{Y})$ as
  \begin{align}
    \label{eq:19}
    h(\mathscr{Y})\approx-\frac{1}{M}\sum_{k=1}^M\int r(y, \vect{L}_{N,k})\log r(y, \vect{L}_{N,k}) \; dy.
  \end{align}
  When $M$ is very large, histogram method can be used to describe
  $\{L_{N,k}\}$ and reduce the computational load.
\end{enumerate}

% \begin{align}
%   \nonumber %\label{eq:35}
%   \begin{split}
%     L_n^{(i)}&=\log\frac{P_{X_n|Y_1^n}(X^{(i)}|Y_1^n)}{P_{X_n|Y_1^n}(X^{(0)}|Y_1^n)},
%     \quad i=1,2,\cdots, m-1 \\
%     L_n^{(i)}&=F^{(i)}(\vect{L}_{n-1})=F^{(i)}(L_{n-1}^{(1)}, L_{n-1}^{(2)},\cdots, L_{n-1}^{(m-1)})
%   \end{split}
% \end{align}
% \begin{figure*}[!t]
%   \begin{align}
%      \begin{split}
%        \label{eq:7}
%         L_{n+1}^{(i)}&=\log\frac{P_{X_{n+1}|Y_1^{n+1}}(X^{(i)}|Y_1^{n+1})}{P_{X_{n+1}|Y_1^{n+1}}(X^{(0)}|Y_1^{n+1})}\\
%     &=\log\frac{\sum_{k=0}^{m-1}
%       P_{Y_{n+1}|X_{n+1}}(Y_{n+1}|X^{(i)})P_{X_2|X_1}(X^{(k)}|X^{(i)})P_{X_n|Y_1^n}(X^{(i)}|Y_1^n)}{\sum_{k=0}^{m-1}
%       P_{Y_{n+1}|X_{n+1}}(Y_{n+1}|X^{(0)})P_{X_2|X_1}(X^{(k)}|X^{(0)})P_{X_n|Y_1^n}(X^{(0)}|Y_1^n)}\\
%     &=\log\frac{P_{Y_{n+1}|X_{n+1}}(Y_{n+1}|X^{(i)})}{P_{Y_{n+1}|X_{n+1}}(Y_{n+1}|X^{(0)})}
%     +\log\frac{\sum_{k=0}^{m-1}P_{X_2|X_1}(X^{(k)}|X^{(i)})\exp(L_n^{(k)} )
%     }{\sum_{k=0}^{m-1}P_{X_2|X_1}(X^{(k)}|X^{(i)})\exp(L_n^{(k)} )}
%   \end{split}
% \end{align}
% \end{figure*}
% where $m$ is the number of the states of Markov Chain. Then
% $P_{Y_{n+1}|Y_1^n}(y|Y_1^n)$ can be calculate by $L^{(i)}_n$,
% $P_{Y|X}$ and $P_{X_{n+1}|X_n}$. And entropy rate can be
% calculated from
% \begin{align}
%   \label{eq:10}
%   P_{X_{n}|Y_1^n}(X^{(i)}|Y_1^n,
%   \mathbf{L}_n)=\frac{\exp(L^{(i)}_n)}{1+\sum_{i=1}^{m-1}\exp(L^{(i)}_n)}.
% \end{align}

\subsection{Numerical Results} \label{sec:NumRes}
\subsubsection{Idealized duty cycle constraint $(q, 0)$}
One implication of Theorem~\ref{thm:Main} is that directly computing
the capacity-achieving input distribution requires solving an
optimization problem with infinite variables which is prohibitive.
Assuming any upper bound on the number of probability mass points,
however, a numerical optimization over the mutual information can
yield a suboptimal input distribution and a lower bound on the channel
capacity. As we increase the number of mass points, the lower bound
can be further refined. We take this approach to numerically compute a
good approximation of the channel capacity by optimizing over a
sufficient number of probability mass points.

Given the duty cycle and power constraints, we first numerically
optimize the mutual information by a 3-point input distribution
(including a mass at 0), then increase the number of probability mass
points by 2 at a time to improve the mutual information, until the
improvement is less than $10^{-3}$.

\begin{figure}[t]
  \centering
  \includegraphics[width=\columnwidth]{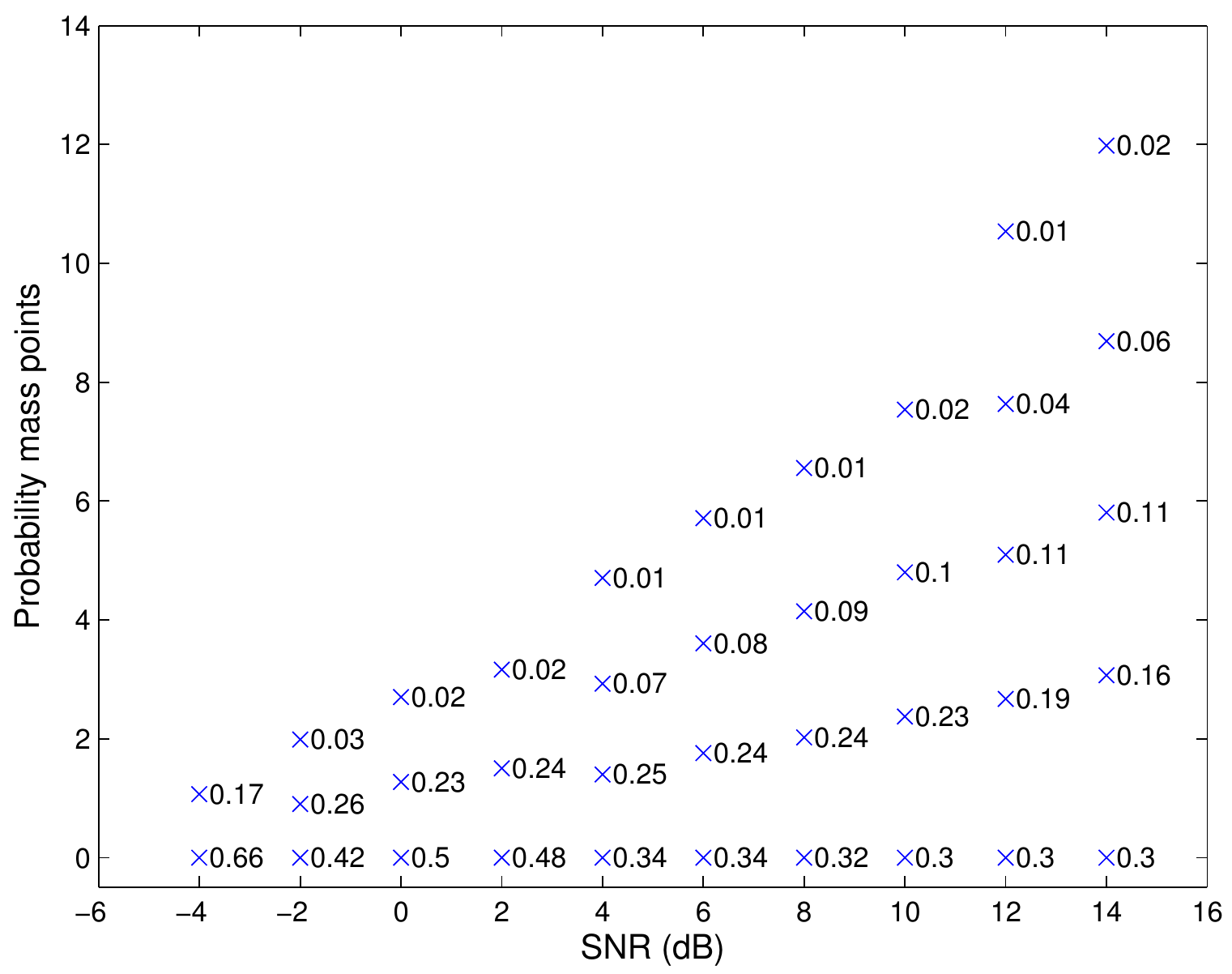}
  \caption{Suboptimal input distribution for $P(X=0)\geq q=0.3$.}
  \label{fig:q3}
\end{figure}

First consider the case that the duty cycle is no greater than $70\%$,
i.e., $P(X=0)\geq q=0.3$. For different SNRs, the mass points of the
near-optimal input distribution with finite support along with the
corresponding probability masses are shown in Fig.~\ref{fig:q3}. Due
to symmetry, only the positive half of the input distribution is
plotted. We can see that as the SNR increases, more masses are put on
higher-amplitude points, whereas the probability mass at zero achieves
its lower bound $0.3$ eventually.

\begin{figure}[t]
  \centering
  \includegraphics[width=\columnwidth]{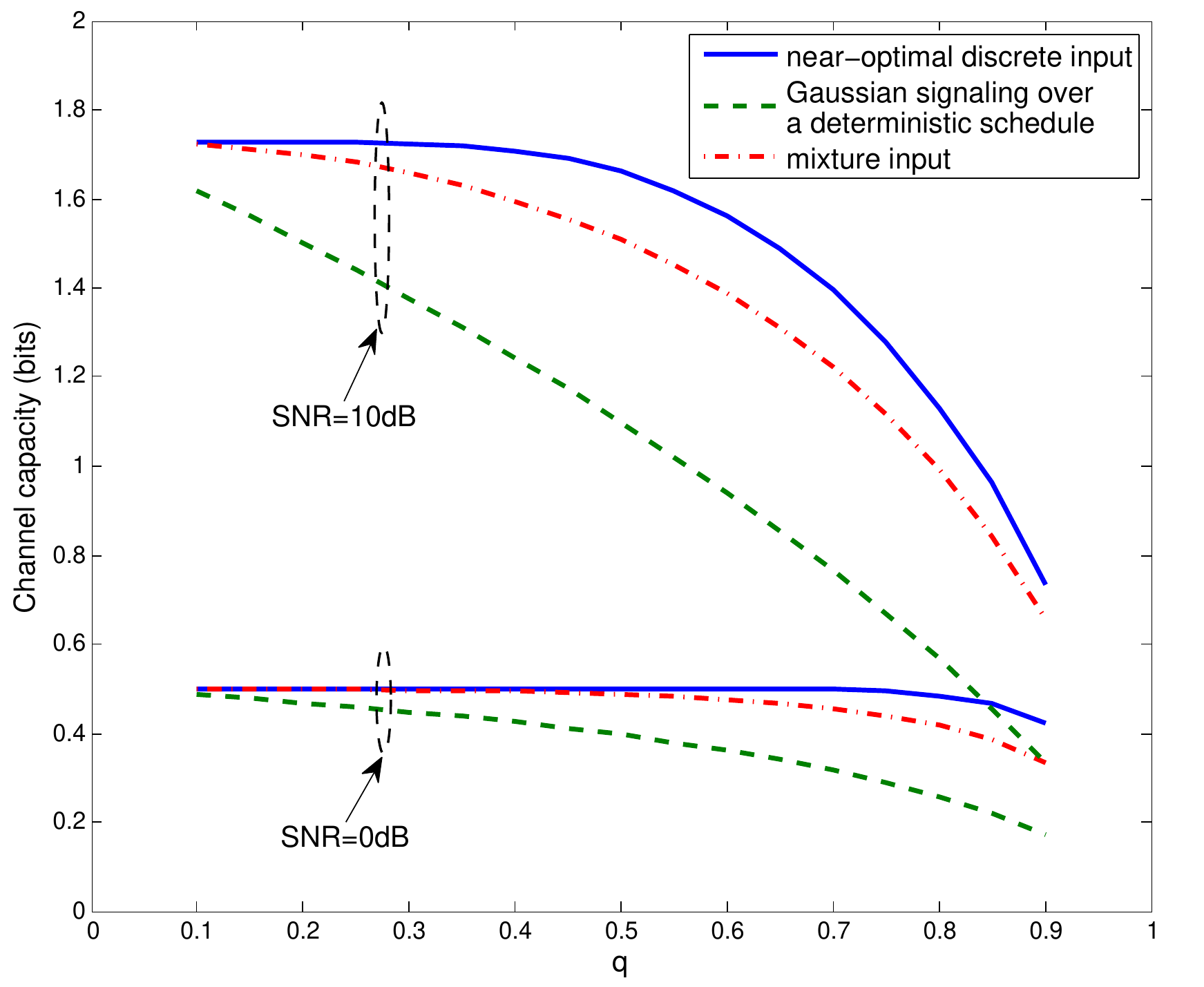}
  \caption{Achievable rates under duty cycle constraint for 0 dB and 10 dB
    SNRs.}
  \label{fig:snr0_10}
\end{figure}

In Fig~\ref{fig:snr0_10}, we compare the rate achieved by the
near-optimal input distribution and the rate achieved by a
conventional scheme using Gaussian signaling over a deterministic
schedule, which is $(1-q)$ times the Gaussian channel capacity without
duty cycle constraint. It is shown in the figure that there is
substantial gain for both 0~dB and 10~dB SNRs by using discrete input
over Gaussian signaling with a deterministic schedule. For example,
when the SNR is 10~dB, given the duty cycle is no more than 50\%, the
discrete input distribution achieves 50\% higher rate. Hence departing
from the usual paradigm of intermittent packet transmissions may yield
significant gains.

We also plot in Fig~\ref{fig:snr0_10} the achievable rate by a superposition coding, where the input distribution is a mixture of Gaussian and a point mass at 0. We first decode the support of the input to find out the positions of nonzero symbols, and then the Gaussian codeword conditioned on the support. It is shown in the figure that the near-optimal discrete input achieves higher rate compared with the mixture input.

\subsubsection{Realistic duty cycle constraint $(q,c)$}

In this subsection the numerical results of lower bound of capacity
and suboptimal distribution are provided based on the results in
Section \ref{sec:lower-bound-capacity} and \ref{sec:numer-comp-entr}.

We first seek a discrete Markov chain with finite alphabet that
maximizes the objective $L(\mu)$ defined in~\eqref{eq:21n}.  Once the
optimal Markov distribution $\mu^*$ is determined, we compute the
achievable rate $I(\mu^*)$ according to~\eqref{eq:17n}.

% \begin{comment}
% \begin{align}
%   \label{eq:12}
%   \begin{split}
%     &\;P_{X_{n+1}|Y_1^n}(X^{(i)}|Y_1^n, \mathbf{L}_n)\\
%     &=\sum_{k=0}^{m-1}P_{X_{n+1}|X_n}(X^{(i)}|X^{(k)})P_{X_{n}|Y_1^n}(X^{(k)}|Y_1^n, \mathbf{L}_n)
%   \end{split}
% \end{align}
% \begin{align}
%   \label{eq:18}
%   \begin{split}
%     p(y|\mathbf{L}_{n})&=P_{Y_{n+1}|Y_1^n}(y|Y_1^n, \mathbf{L}_n)\\
%     &=\sum_{i=0}^{m-1}P_{Y_{n+1}|X_{n+1}}(y|X^{(i)}_{n+1})P_{X_{n+1}|Y_1^n}(X^{(i)}|Y_1^n)
%   \end{split}
% \end{align}
% \begin{align}
%   \label{eq:24}
%   \begin{split}
%     h(\mathscr{Y})&=\lim_{n\to\infty}h(Y_{n+1}|Y_1^n) \\
%     &=\lim_{n\to\infty}-\int\int p(y|\mathbf{L}_{n})\log
%     p(y|\mathbf{L}_{n})dy dP(\mathbf{L}_n)\\
%   \end{split}
% \end{align}
% \end{comment}
\begin{table}[!t]
  \centering
  \begin{tabular}{rr|rrrrr}
 &   & \multicolumn{5}{c}{\raisebox{1ex}[0pt]{$X_2$}}\\
       &      & 0.0000 &   3.9281 &  -3.9281 & 7.1398 & -7.1398  \\\hline
 &     0.0000 & 0.8342 &   0.0605 & 0.0605 & 0.0224 &      0.0224 \\
 &     3.9281 & 0.4923 &   0.1852 & 0.1852 & 0.0687 &      0.0687 \\
$X_1$& -3.9281 & 0.4923 &   0.1852 & 0.1852 & 0.0687 &      0.0687 \\
    & 7.1398 & 0.4923 &   0.1852  &  0.1852 &  0.0687 &     0.0687 \\
    & -7.1398 & 0.4923 &   0.1852  &   0.1852 & 0.0687 &     0.0687
 \\\hline
 & $P_X$ &  0.7481 &   0.0919 &   0.0919  &  0.0341 &     0.0341
  \end{tabular}
  \caption{$P_{X_2|X_1}$ and $P_X$ for $q=0.5$, $c=1.0$, $\mathrm{SNR=8dB}$.}  \label{tab:1}
%  \vspace{-8ex}
\end{table}
\begin{figure}[!t]
  \centering
  \includegraphics[width=\columnwidth]{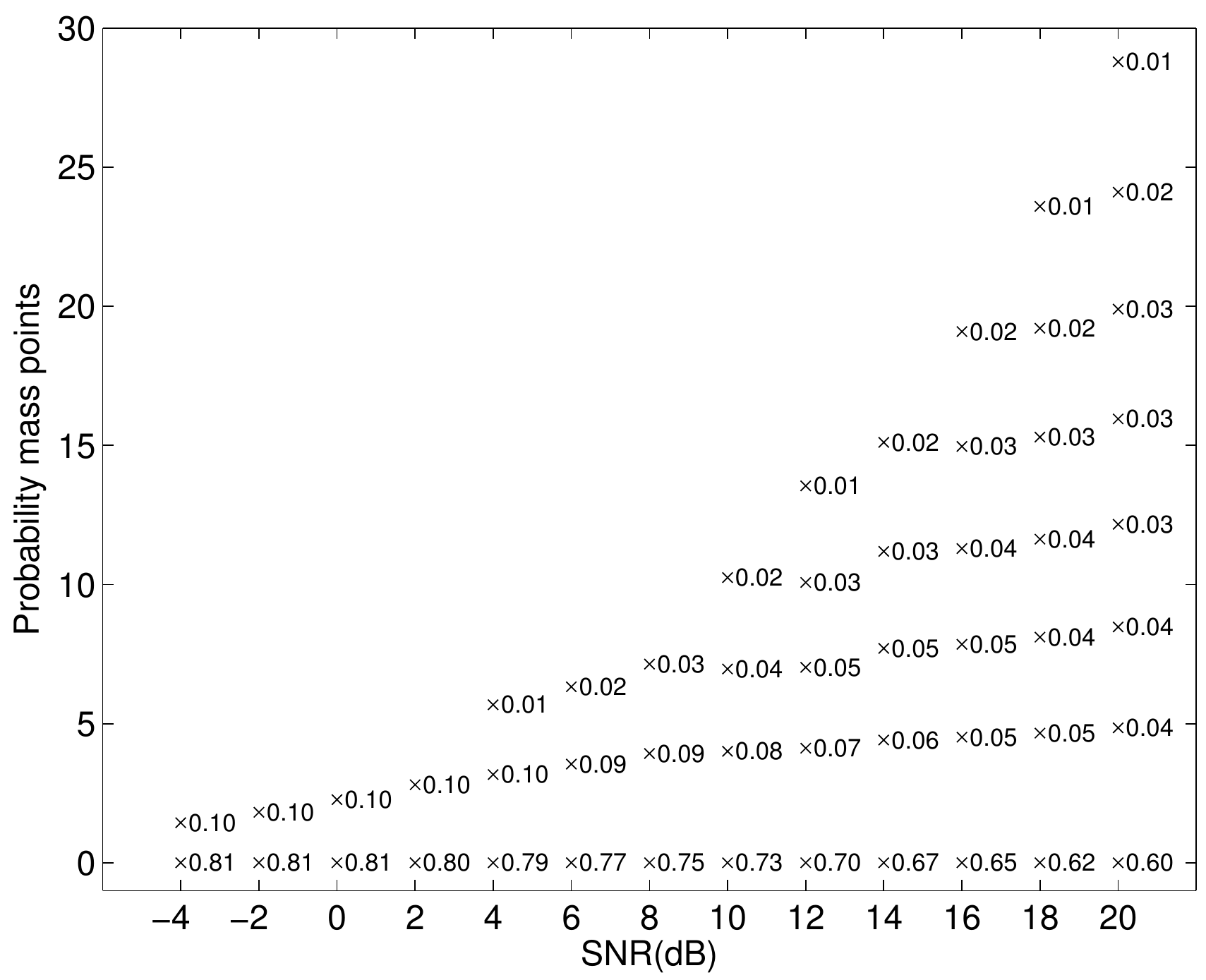}
  \caption{The marginal distribution of the stationary Markov input.
    Duty cycle $\le 0.5$, transition cost $c=1.0$.}\label{fig:1}
\end{figure}
\begin{figure}[!t]
  \centering
  \includegraphics[width=\columnwidth]{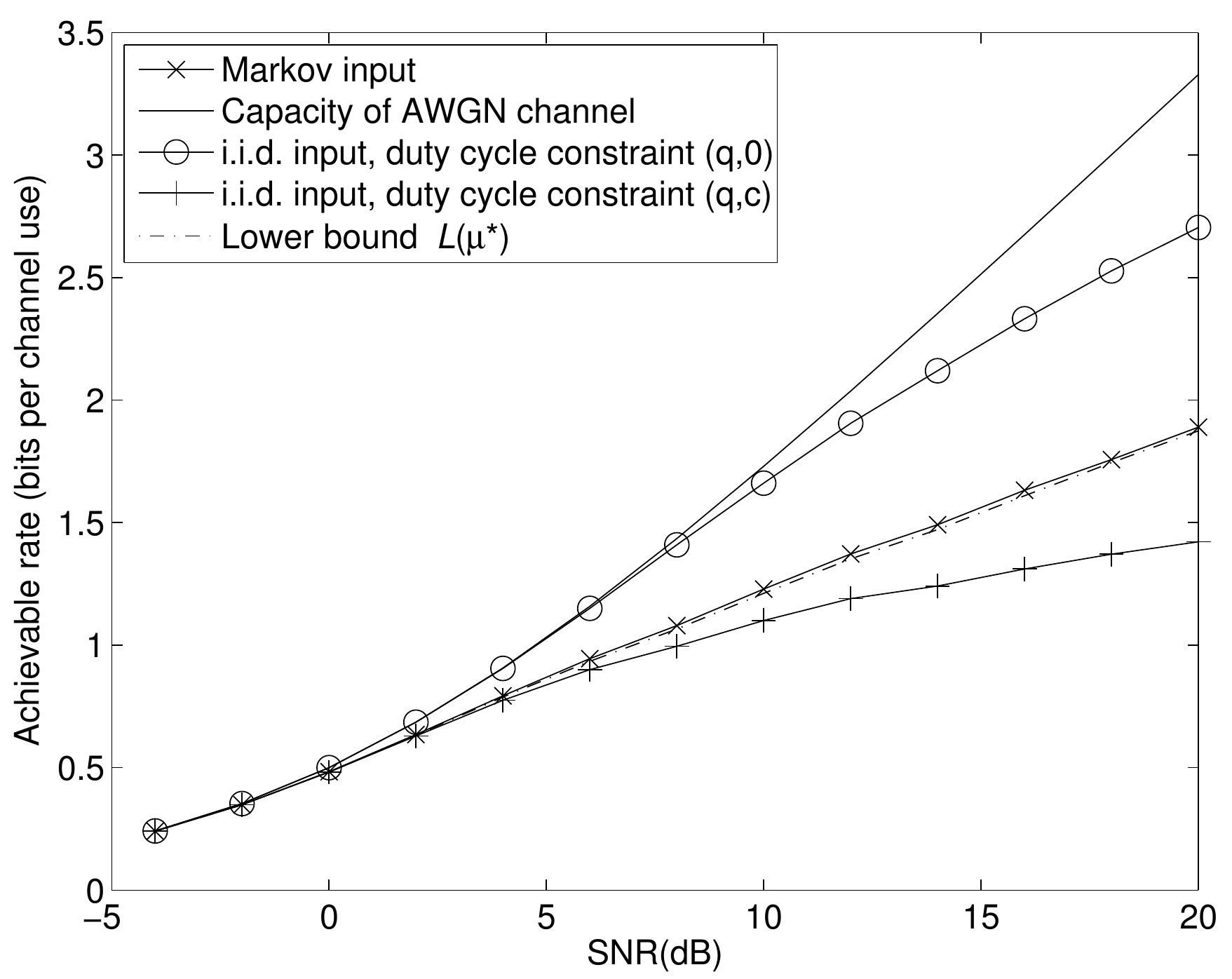}
  \caption{The achievable rate vs. the SNR. Duty cycle $\le 0.5$, transition cost $c=1.0$.}\label{fig:2}
  \vspace{-1ex}
\end{figure}
\begin{figure}[!t]
  \centering
  \includegraphics[width=\columnwidth]{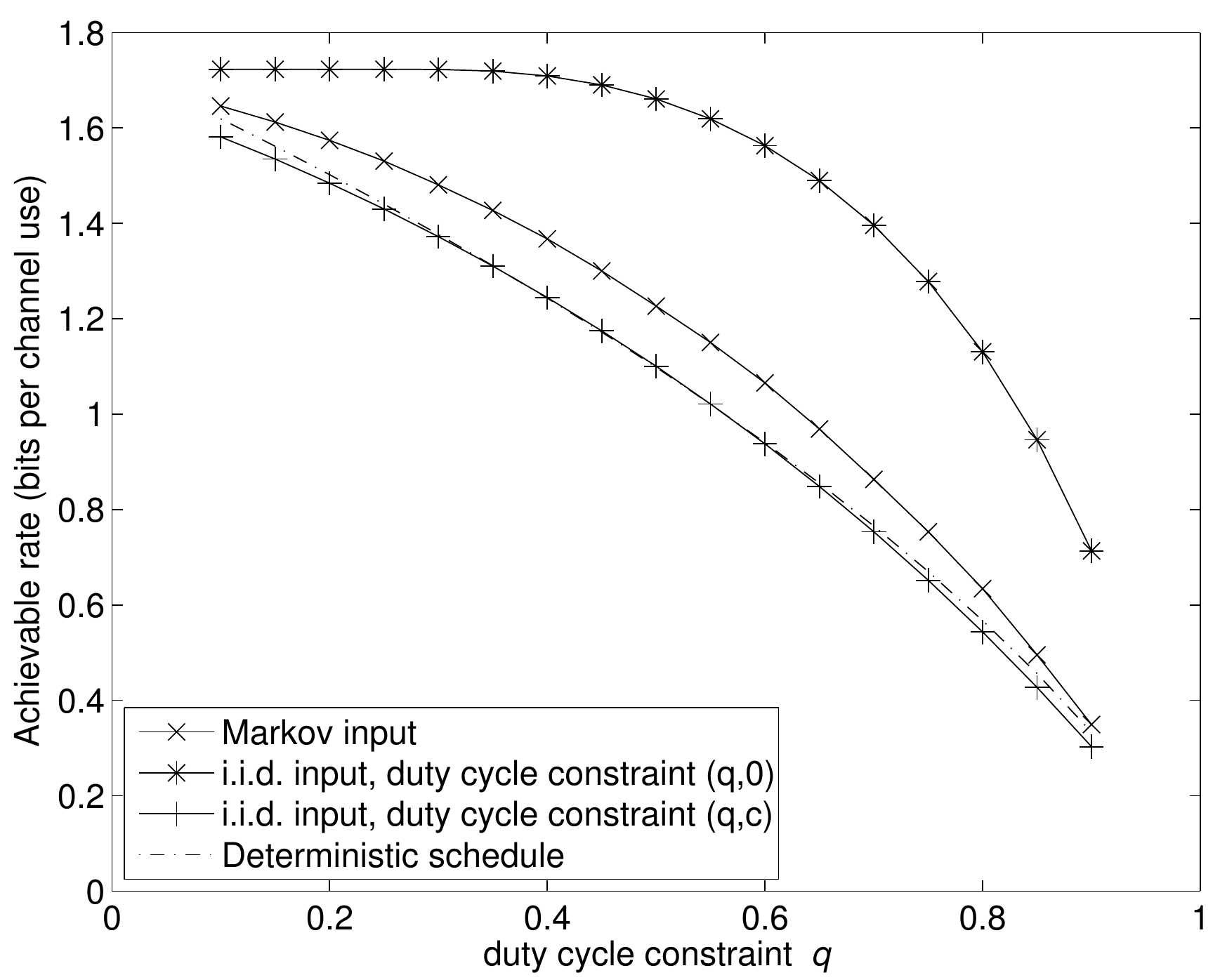}
  \caption{The achievable rate vs. the duty cycle,
    SNR = 10 dB and transition cost $c=1.0$.} \label{fig:3}
\end{figure}
\begin{figure}[!t]
  \centering
  \includegraphics[width=\columnwidth]{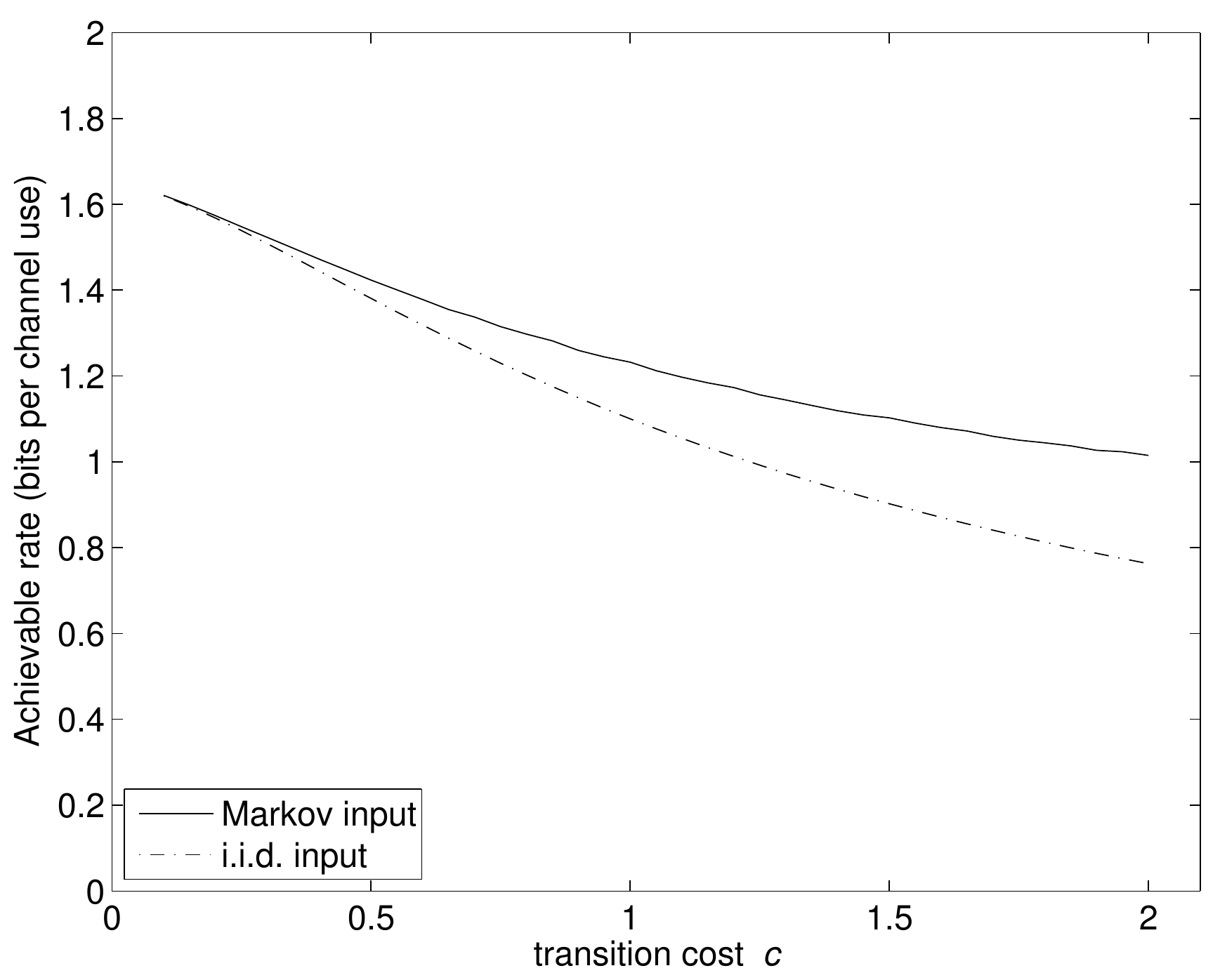}
  \caption{The achievable rate vs. transition cost, with SNR = 10 dB
    and $q=0.5$.}\label{fig:4}
\end{figure}

In this paper $\mu^*=(\mathcal{X},\alpha, \beta, P_X)$ is used to
approximate the optimum distribution $\mu_0$ through the
maximizing $L(\cdot)$. It is obvious that the optimized $\mu^*$
is symmetric about 0. Table \ref{tab:1} is the transition
probability matrix $P_{X_2|X_1}$ and stationary probability $P_X$ for
$q=0.5$, $c=1.0$ and SNR = 8 dB.  The symmetry of the transition
probability matrix is evident, as conditioned on that two consecutive
symbols are nonzero, they are independent.

Fig.~\ref{fig:1} shows the stationary (marginal)
distribution for suboptimal Markov input. In order to compensate the
transition cost, additional fraction of zero symbol should be
transmitted, $P_X(0)>q$.
As the SNR increases, more and more weights are put on distant
constellation points, where less and less weights are put on the zero letter.

In Fig.~\ref{fig:2}, the rates achieved by various optimized input
distributions are plotted against the SNR.  The rate achieved by
the optimized Markov input is larger than that of suboptimal i.i.d.~input
calculated by formula (\ref{eq:16n}) with duty cycle
constraint $(q,c)$. The lower bound $L(\mu)$ is quite tight and can
be regarded as a good approximation of mutual information of first-order
Markov inputs.

Figs.~\ref{fig:3} and~\ref{fig:4} demonstrate the sensitivity of the
achievable rates to the duty cycle parameter $q$ and the transition
cost $c$, respectively.  The performance of Markov inputs is superior
to i.i.d.~inputs as well as Gaussian signaling with deterministic
schedule. Fig \ref{fig:3} shows that the performance of i.i.d. input
is similar to the deterministic schedule, which implies that different
from the case under the idealized duty cycle constraint, i.i.d. input
is not a good choice under the realistic duty cycle constraint.

\section{Concluding Remarks}
\label{sec:conclusion}
In this paper we have studied the impact of duty cycle constraint on
the capacity of AWGN channels. Under the idealize duty cycle
constraint, the optimal distribution has an infinite number of
probability mass points in a bounded interval. This allows efficient
numerical optimization of the input distribution.  Under the realistic
duty cycle constraint, the capacity-achieving input is hard to compute.
We develop techniques for computing a near-optimal input distribution.
This input takes the form of a discrete first-order Markov process,
which matches the ``Markov'' nature of the duty cycle constraint.  The
numerical results show that under the duty cycle constraint, departing
from the usual paradigm of intermittent packet transmissions may yield
substantial gain.

\section*{Acknowledgement}

D. Guo and L. Zhang would like to thank Raymond Yeung and Shuo-Yen Robert Li
for hosting them in the Institute of Network Coding at the Chinese
University of Hong Kong during the production of the first draft of
this paper.  The authors would also like to thank Terence Chan for
sharing the code for the numerical results in~\cite{ChaHra05IT} and
Yihong Wu and Sergio Verd\'u for their helpful comments.

\bibliographystyle{ieeetr}
\bibliography{IEEEabrv,str_def,DutyCycle12IT}

\end{document}